# A fast and accurate semi-analytical method to determine the thermal response of bore fields

Enzo Zanchini [a,*], Francesco Zanchini [b]

[a] Department of Industrial Engineering, Viale del Risorgimento 2, University of Bologna, Bologna, Italy
[*]Corresponding author - Email: enzo.zanchini@unibo.it

[b] Student of the Department of Physics and Astronomy, Viale Carlo Berti Pichat 6/2, University of Bologna, Bologna, Italy
Email: francesco.zanchini4@studio.unibo.it

## Abstract

The design and the simulation of a borehole-heat-exchanger (BHE) field is usually performed by simplified methods that yield either an overestimation or an underestimation of the thermal response. The methods employing the assumption of a uniform heat rate per unit BHE length overestimate the thermal response, while those employing the assumption of a uniform temperature of the external surface of the BHEs underestimate it. An accurate semi-analytical method to determine the *g-function* of a bore field with the real condition of BHEs fed in parallel with equal inlet temperature was developed by Cimmino (Int J Heat Mass Tran 91, 2015). An alternative semi-analytical method that yields the same results is presented in this paper. The method is implemented in a C++ program, available at the open-source online data repository of the University of Bologna. Thanks to several optimizations, the program yields a very accurate thermal response of bore fields of any shape with an extremely short computation time. The program is employed to analyze the inaccuracies caused by the assumptions of uniform heat rate and uniform surface temperature of the BHEs. It is also used to illustrate the low performance of the central BHEs in large and compact bore fields, and to show how the bore field can be optimized for a given plot of land and a fixed total length of the field.

**Nomenclature**

| | |
|---|---|
| 2U102 1.0 | Type of double U-tube BHE, defined in Table 1 |
| 2U102 1.6 | Type of double U-tube BHE, defined in Table 1 |
| 2U85 1.0 | Type of double U-tube BHE, defined in Table 1 |
| 2U85 1.6 | Type of double U-tube BHE, defined in Table 1 |
| $A_{mn}$ | Dimensionless $N \times N$ matrix, defined in Eq. (28) |
| $A_{mp}$ | Dimensionless $P \times P$ matrix, defined in Eq. (40) |
| $a$ | Dimensionless heat load per unit length, $= \dot{q}_l / \dot{q}_{l0}$ |
| $\bar{a}$ | Dimensionless heat load per unit length averaged along a BHE |
| $B$ | Distance between BHEs $(\mathrm{m})$ |
| BHE | Borehole Heat Exchanger |
| $B_x$, $B_y$ | Distance between BHE columns, distance between the BHE rows $(\mathrm{m})$ |
| $b_m$ | Dimensionless vector, defined in Eqs. (29) and (41) |
| $c_{p\mathrm{f}}$ | Specific heat capacity at constant pressure of the fluid $(\mathrm{J\ kg^{-1}K^{-1}})$ |
| $C$ | Dimensionless parameter, defined in Eq. (54) |
| $D$ | Buried depth $(\mathrm{m})$ |
| $D^*$ | Dimensionless buried depth, $= D/H_\mathrm{b}$ |
| $d^*_{mn}$ | Dimensionless horizontal distance between segments $m$ and $n$ |
| erf | Error function, defined in Eq. (13) |
| $f$ | Reduction factor in the number of operations to solve the linear systems |
| FLS | Finite line-source |
| $g$ | *g-function* |
| $h$ | Convection coefficient $(\mathrm{W\ m^{-2}K^{-1}})$ |
| $H^*$ | Dimensionless length of each borehole segment, |
| $H_\mathrm{b}$ | Length of each BHE $(\mathrm{m})$ |
| $h_{mn}$ | Segment-to-segment finite line-source solution |
| $h^+_{mn}$, $h^-_{mn}$ | Partial contributions to $h_{mn}$, defined in Eqs. (31, 32) |
| Ierf | Integral error function, defined in Eq. (12) |
| $k_\mathrm{f}$ | Thermal conductivity of the fluid $(\mathrm{W\ m^{-1}K^{-1}})$ |
| $k_\mathrm{g}$ | Thermal conductivity of the ground $(\mathrm{W\ m^{-1}K^{-1}})$ |
| $k_\mathrm{gt}$ | Thermal conductivity of the grout $(\mathrm{W\ m^{-1}K^{-1}})$ |
| $k_\mathrm{p}$ | Thermal conductivity of the pipe $(\mathrm{W\ m^{-1}K^{-1}})$ |
| $L$ | Total length of the bore field $(\mathrm{m})$ |
| $\ell$ | Length (m) |
| $\ell^*$ | Dimensionless length, $= \ell/H_\mathrm{b}$ |
| $\dot{m}$ | Total mass flow rate (kg $\mathrm{s}^{-1}$) |
| $N$ | Total number of segments |
| $N_\mathrm{b}$ | Number of boreholes |
| $N_\mathrm{s}$ | Number of segments in which each BHE is divided |
| $P$ | Number of sets of equivalent segments |
| $\dot{Q}$ | Total heat rate supplied to the bore field (W) |
| $\dot{q}_l$ | Mean heat rate per unit length (W $\mathrm{m}^{-1}$) |
| $R_{11}$, $R_{12}$, $R_{13}$ | Thermal resistances, defined in Eqs. (48, 49, 50) (m K $\mathrm{W}^{-1}$) |
| $R_\mathrm{a}$ | Total thermal resistance between the pipes (m K $\mathrm{W}^{-1}$) |
| $R_\mathrm{b}$ | Thermal resistance of a BHE cross section (m K $\mathrm{W}^{-1}$) |
| $R_\mathrm{b3D}$ | 3D borehole thermal resistance (m K $\mathrm{W}^{-1}$) |
| $R^*_\mathrm{b3D}$ | Dimensionless 3D borehole thermal resistance |
| $R_\mathrm{beff}$ | Effective borehole thermal resistance (m K $\mathrm{W}^{-1}$) |
| $R_\mathrm{p}$ | Thermal resistance of the pipe (m K $\mathrm{W}^{-1}$) |
| $r_\mathrm{b}$ | Radius of the BHE (m) |
| $r_\mathrm{e}$ | External radius of the pipe (m) |
| $r_\mathrm{i}$ | Internal radius of the pipe (m) |
| $s$ | Half shank spacing (m) |
| $S$ | Dimensionless parameter, defined in Eq. (55) |
| $T$ | Temperature (°C) |
| $T^*$ | Dimensionless temperature, defined in Eq. (6) |
| $t$ | Time (s) |
| $t^*$ | Dimensionless time, defined in Eq. (7) |
| $t_\mathrm{hours}$ | Time in hours $= t/1$ hour |
| $t_\mathrm{years}$ | Time in years $= t/1$ year |
| $t^*_K$ | Last dimensionless time instant considered |
| $T_\mathrm{b}$ | Mean temperature of the BHE external surface (°C) |
| $T^*_\mathrm{b}$ | Dimensionless mean temperature of the external surface of the BHEs |
| $T_\mathrm{f}$ | Mean temperature of the fluid in the BHEs (°C) |
| $T^*_\mathrm{f}$ | Dimensionless mean temperature of the fluid |
| $T_\mathrm{g}$ | Undisturbed ground temperature (°C) |
| $T_{\text{in}}$ | Inlet fluid temperature (°C) |
| $T_{\text{out}}$ | Outlet fluid temperature (°C) |
| $U$ | Heaviside unit-step function |
| U94 1.0 | Type of single U-tube BHE, defined in Table 1 |
| U94 1.6 | Type of single U-tube BHE, defined in Table 1 |
| U54 1.0 | Type of single U-tube BHE, defined in Table 1 |
| U54 1.6 | Type of single U-tube BHEs (see Table 1) |
| $\dot{V}$ | Volume flow rate supplied to each BHE ($\text{m}^3\ \text{s}^{-1}$) |
| $w$ | Collection of segments |
| $x, y$ | Horizontal coordinates (m) |
| $z$ | Distance from the BHE top (m) |
| $z^*$ | Dimensionless distance from the BHE top |

Subscripts

| | |
|---|---|
| 0 | Reference value, constant value, of segment number 0 |
| $i$ | Referred to the *i*-th time instant |
| $k$ | Referred to the *k*-th time instant |
| $m$ | Of the *m*-th segment |
| $n$ | Of the *n*-th segment |
| $mn$ | Referred to the *m*-th and the *n*-th segment |
| $p$ | Of the *p*-th set of equivalent segments |
| $uv$ | Of the BHE located at the *u*-th row and the *v*-th column |

Greek symbols

| | |
|---|---|
| $\alpha_{\text{g}}$ | Thermal diffusivity of the ground ($\text{m}^2\text{s}^{-1}$) |
| $\delta_{mn}$ | Kronecker delta function, defined in Eq. (25) |
| $\eta$ | Dimensionless parameter, defined in Eq. (47) |
| $\Theta$ | Dimensionless bulk fluid temperature, defined in Eq. (53) |
| $\Theta_{\text{d}}$ | Dimensionless bulk fluid temperature in the downward flow |
| $\Theta_{\text{u}}$ | Dimensionless bulk fluid temperature in the upward flow |
| $\mu_{\text{f}}$ | Dynamic viscosity of the fluid (Pa s) |
| $\rho_{\text{f}}$ | Density of the fluid ($\text{kg m}^{-3}$) |
| $\sigma$ | Dimensionless parameter, defined in Eq. (45) |
| $\varphi$ | Dimensionless coefficient, defined in Eq. (3) |

## 1. Introduction

Ground-coupled Heat Pumps are the most efficient systems for heating and cooling buildings [1]. A Ground-coupled Heat Pump is usually equipped with a field of vertical ground heat exchangers, called Borehole Heat Exchangers (BHEs). In most applications, BHEs are composed of either a single U-tube or a double U-tube in high density polyethylene, surrounded by a sealing grout.

The design and the simulation of a bore field is based on dimensionless functions called thermal response factors. A thermal response factor yields the time evolution of either the mean temperature of the external surface of the BHEs, $T_b$, or the mean temperature of the fluid in the BHEs, $T_f$, caused by a time constant heat rate supplied to the BHEs. The case of a time dependent heat rate can then be studied by applying the superposition of the effects in time.

The most widely employed thermal response factors yield the time evolution of $T_b$, and are called *g-functions*. The concept of *g-function* was first introduced by Eskilson [2], who numerically determined the *g-functions* of bore fields with the boundary condition of uniform temperature of the external surface of the bore field and time constant total heat rate supplied to the bore field.

A simpler boundary condition is that of uniform and time constant heat rate. Under this condition, some authors determined analytical expressions of the temperature averaged along the BHE depth, at a radial distance $r$ from the BHE axis, by sketching the BHE as a finite line-source (FLS). An analytical expression of the temperature averaged along the BHE depth, with this boundary condition, was first found by Zeng et al. [3], for a BHE with its top at the ground surface. Simpler expressions were proposed by Lamarche and Beauchamp [4] and by Bandos et al. [5]. An analytical expression of the temperature averaged along the BHE depth was then obtained by Claesson and Javed [6] in the more general case of a BHE with its top buried at a depth $D$. Under the assumption of uniform and constant heat rate, the *g-function* of a BHE field can be easily evaluated by the superposition of the effects of the single BHEs.

In real applications, however, the BHEs of a field are fed in parallel with the same inlet temperature, and the heat load is neither the same for all BHEs, nor uniform along each BHE. Therefore, the condition of uniform temperature of the external surface of the bore field, employed by Eskilson, seems closer to the real condition.

Cimmino and Bernier [7] found a semi-analytical method to generate the *g-functions* of bore fields with Eskilson's boundary condition, and proved that the *g-functions* determined by the FLS solution with uniform heat load overestimate the thermal response of bore fields with respect to the *g-functions* obtained with Eskilson's boundary condition. Lamarche [8] and Cimmino [9] proposed improvements of the method developed in Ref. [7], aimed at reducing the computation time. Monzó et al. [10] and Naldi and Zanchini [11] developed finite-element numerical methods to obtain *g-* *functions* with Eskilson's boundary condition, and confirmed the overestimation of the thermal response yielded by the FLS solution with uniform heat load.

In fact, while the fluid temperature averaged between the pipes is almost independent of the vertical coordinate, $z$, the surface temperature of the BHEs varies with $z$, due to the combined effect of the borehole thermal resistance and the non-uniform heat load per unit length, so that even Eskilson's boundary condition does not correspond to the real working condition.

Monzó et al. [12] refined the numerical method of Ref. [10] to take into account the effects of the borehole thermal resistance. Cimmino [13] developed a semi-analytical method to obtain *g-functions* with the boundary condition of equal inlet fluid temperature for all the boreholes in the field, and showed that the boundary conditions of uniform surface temperature and uniform heat rate yield accurate *g-functions*, respectively, only in the limits of very low and very high borehole thermal resistance. The same author then proposed an extension of the method, which allows the simulation of bore fields with series and parallel-connected boreholes [14].

When the design or the simulation of a bore field is performed through the *g-function*, it is necessary to derive the time evolution of the outlet fluid temperature, $T_{\mathrm{out}}$, from that of the mean temperature of the external surface, $T_{\mathrm{b}}$. Usually, the mean fluid temperature, $T_{\mathrm{f}}$, is first determined by the relation

$$T_{\mathrm{f}} = T_{\mathrm{b}} + \dot{q}_l R_{\mathrm{b}}, \tag{1}$$

where $\dot{q}_l$ is the mean heat rate per unit length supplied to the bore field, and $R_{\mathrm{b}}$ is the steady-state thermal resistance of a BHE cross section, evaluated either by an approximate analytical expression of by a 2D numerical simulation. Then, the outlet fluid temperature, $T_{\mathrm{out}}$, is determined by the relation

$$T_{\mathrm{out}} = T_{\mathrm{f}} - 0.5 \frac{\dot{Q}}{\dot{m} c_{pf}}, \tag{2}$$

where $\dot{Q}$ is the total heat rate supplied to the bore field, $\dot{m}$ is the total mass flow rate, and $c_{pf}$ is the specific heat capacity at constant pressure of the fluid.

Both Eq. (1) and Eq. (2) are inaccurate. The inaccuracy of Eq. (2) is due to the fact that the bulk temperature of the fluid is not a linear function of the distance from the inlet. This inaccuracy can be easily corrected by employing the relation between $T_{\mathrm{out}}$ and $T_{\mathrm{f}}$ recommended in Ref. [15], namely

$$T_{\mathrm{out}} = T_{\mathrm{f}} - \left(0.5 - \varphi \frac{\dot{V}}{\dot{V}_0}\right) \frac{\dot{Q}}{\dot{m} c_{pf}}\,. \tag{3}$$

In Eq. (3), $\dot{V}$ is the volume flow rate supplied to each BHE, $\dot{V}_0$ is a reference volume flow rate, equal to 12 liters per minute, and $\varphi$ is a dimensionless coefficient to be determined by the correlations given in Ref. [15] for single U-tube BHEs and in Ref. [16] for double U-tube BHEs.

The inaccuracies in Eq. (1) are more difficult to correct. In the short term, i.e., when the heat transfer within the BHE is still transient, the difference $T_{\mathrm{f}} - T_{\mathrm{b}}$ is much lower than $\dot{q}_l R_{\mathrm{b}}$. This difference can be expressed by defining a time dependent 3D thermal resistance,

$$R_{\mathrm{b3D}} = \frac{T_{\mathrm{f}} - T_{\mathrm{b}}}{\dot{q}_l}, \tag{4}$$

where $T_{\mathrm{f}}$ is evaluated as an integral over the region of space occupied by the fluid, divided by the volume of that region, and $T_{\mathrm{b}}$ is evaluated as an integral over the BHE external surface, divided by the area of that surface. In the short term, $R_{\mathrm{b3D}}$ is much lower than $R_{\mathrm{b}}$. In the long term, $R_{\mathrm{b3D}}$ reaches a quasi-stationary value higher than $R_{\mathrm{b}}$, due to the thermal short circuiting between the legs of the BHEs. The time evolutions of the ratio $R_{\mathrm{b3D}}/R_{\mathrm{b}}$ for some typical single U-tube and double U-tube BHEs, of length 100 m, are illustrated in Ref. [17].

In order to obtain an accurate time evolution of $T_{\mathrm{out}}$ in the short term, some authors developed lumped-parameters numerical models based on the electric analogy, where the BHE and the ground are represented by a grid of thermal resistances and heat capacities [18, 19, 20, 21]. These models are suitable for reproducing the transient behavior of the BHE in the short term, but become inaccurate in the medium and long term. Other authors [22, 23, 24, 25] developed 1D analytical models that yield an accurate estimation of the time evolution of $T_{\mathrm{f}}$ in the short term, but become inaccurate in the long term, due to neglecting the effects of the heat conduction in the axial direction.

Man et al. [26] and Li et al. [27] developed analytical BHE models that consider the heat conduction in the vertical direction and yield accurately the time evolution of $T_{\mathrm{f}}$ both in the short and in the long term. In these models, which will be called analytical conduction models, the fluid is replaced by a heat generating surface [26], or by a heat-generating set of lines [27]. Analytical conduction models can be used for an accurate simulation of a single BHE, but should not be used for the simulation of bore fields by applying the superposition of the effects in space. In fact, they consider a uniform heat rate supplied to the BHEs, and would yield an overestimation of the thermal response [7, 13].

A numerical conduction model suitable for the simulation of bore fields was proposed by Zanchini et al. [28]. The model yields the time evolution of $T_{\mathrm{f}}$ with the condition of uniform fluid temperature, and was applied to provide tables of fluid-to-ground thermal response factors (*ftg-functions*), that give, in dimensionless form, the time evolution of $T_{\mathrm{f}}$ of any single-line bore field with up to 4 BHEs. In order to obtain a precise thermal response factor in the medium and long term, the model should be applied by employing a suitably reduced thermal conductivity of the sealing grout, to take into account the thermal short circuiting between descending and ascending fluid [17]. If applied in this way, the model yields quite accurate results, but requires high RAM and very long computation times when applied to bore fields with several BHEs.

Complete numerical simulation models have also been developed, where the simulation of the heat conduction in the solid materials is coupled with the energy balance along the flow in the pipes [29, 30, 31, 32, 33, 34]. Complete numerical models yield very precise time evolutions of $T_{\mathrm{f}}$ and of $T_{\mathrm{out}}$, but are unsuitable for the long-term simulation of large bore fields.

The best methods for the simulation of bore fields seem to be those proposed by Cimmino and Bernier [7] and by Cimmino [13, 14]. The method proposed in Ref. [7] is a revolutionary contribution to the simulation of bore fields. In this method, each BHE is divided into segments. Each segment is subjected to a uniform linear heat load, and the magnitude of each linear heat load is determined by imposing the conditions of uniform temperature of the external surface of the bore field and time constant total heat load. Important improvements of the method are proposed in Refs. [13, 14], where the condition of uniform surface temperature of the bore field is replaced by that of equal inlet temperature in all the BHEs. The improved methods yield fully accurate *g-functions* of bore fields. In order to obtain accurate predictions of the time evolution of $T_{\mathrm{f}}$ in the short term, these methods need to be completed by a suitable short-term simulation model.

In this paper, we provide a new semi-analytical method for the calculation of *g-functions* and *ftg-functions* of bore fields, which is an alternative to that proposed in Ref. [13]. In this method, each segment is subjected to a uniform time-dependent linear heat rate, and the magnitude of each linear heat rate is determined by imposing the conditions of uniform fluid temperature and time-constant total heat rate for the entire field. A simulation program based on this method was written in C++. The program yields the *g-function*, the *ftg-function*, and the time evolution of the linear heat load of each BHE segment, for any bore field composed of boreholes with the same length fed in parallel. The optimizations introduced in the evaluation of the FLS solutions for borehole segments and in the solution of the systems of equations allow very high performances in terms of time efficiency.

## 2. Proposed model

Consider a bore field of $N_{\mathrm{b}}$ identical boreholes of length $H_{\mathrm{b}}$, so that the total length of the bore field is $L = N_{\mathrm{b}} H_{\mathrm{b}}$. Assume that the field is subjected to a time constant heat rate $\dot{Q}_0$. Denote by $\dot{q}_{l0}$ the mean heat rate per unit length, namely

$$\dot{q}_{l0} = \frac{\dot{Q}_0}{L} \,. \tag{5}$$

Define the dimensionless temperature as

$$T^* = \frac{k_{\mathrm{g}}}{\dot{q}_{l0}} \left(T - T_g\right), \tag{6}$$

where $T$ is the dimensional temperature, $T_{\mathrm{g}}$ is the undisturbed ground temperature, and $k_{\mathrm{g}}$ is the thermal conductivity of the ground.

Define the dimensionless time as

$$t^* = t \frac{9\alpha_{\mathrm{g}}}{{H_{\mathrm{b}}}^2}, \tag{7}$$

where $t$ is the dimensional time and $\alpha_{\mathrm{g}}$ is the thermal diffusivity of the ground.

Finally, given any length $\ell$, define the corresponding dimensionless length as

$$\ell^* = \frac{\ell}{H_\mathrm{b}}. \qquad (8)$$

Assume that each BHE can be sketched as a finite line-source in a semi-infinite homogeneous solid that represents the ground, and that the temperature of the ground surface is uniform and constant. Assume that either the linear heat load is uniform, or that the temperature of the external surface of the BHEs is uniform. Then, the mean dimensionless temperature of the external surface of the BHEs, $T_\mathrm{b}^*$, is a function of $t^*$ that depends only on the dimensionless parameters that characterize the geometry of the bore field, and is independent of the value of $\dot{q}_{l0}$ and of the thermal properties of the ground. This function is called *g-function*, and is denoted here by $g(t^*)$. In symbols, one has

$$T_\mathrm{b}^*(t^*) = g(t^*). \qquad (9)$$

The assumption of uniform heat load yields an overestimation of $g(t^*)$, while that of uniform surface temperature yields an underestimation with respect to the real working conditions. Therefore, we impose the condition of uniform temperature of the working fluid, by the following procedure.

We divide each BHE into $N_\mathrm{s}$ equal segments, so that the total number of segments is $N = N_\mathrm{b} N_\mathrm{s}$. We denote by $n$ the generic BHE segment, where $n$ ranges from 0 to $N-1$.

While the mean heat load per unit length applied to the bore field, $\dot{q}_{l0}$, is constant in time, the heat load per unit length applied to each segment may depend on time and on the segment considered. We then denote by $\dot{q}_{l,n}(t^*)$ the heat load per unit length applied to the *n*-th segment as a function of time. We also denote by

$$a_n(t^*) = \frac{\dot{q}_{l,n}(t^*)}{\dot{q}_{l0}} \qquad (10)$$

the dimensionless heat load applied to *n*-th segment.

The dimensionless temperature of the external surface of the *m*-th segment, $T_{\mathrm{b},m}^*(t^*)$, increases due to the heat load applied to all the segments, including the *m*-th segment itself. We can then express $T_{\mathrm{b},m}^*(t^*)$ as a sum of contributions:

$$T_{\mathrm{b},m}^*(t^*) = \sum_{n=0}^{N-1} T_{\mathrm{b},mn}^*(t^*), \qquad (11)$$

where $T_{\mathrm{b},mn}^*$ is the contribution to the temperature increase of the *m*-th segment due to the heat load of the *n*-th segment.

Consider now a specific pair of segments, $m$ and $n$. If the dimensionless heat load $a_n(t^*)$ of the segment $n$ has a constant value equal to 1 from the start of the heating, the contribution $T_{\mathrm{b},mn}^*(t^*)$ to the temperature increase of the segment $m$ can be computed analytically, and it is denoted as $h_{mn}(t^*)$ (see Fig. 1).

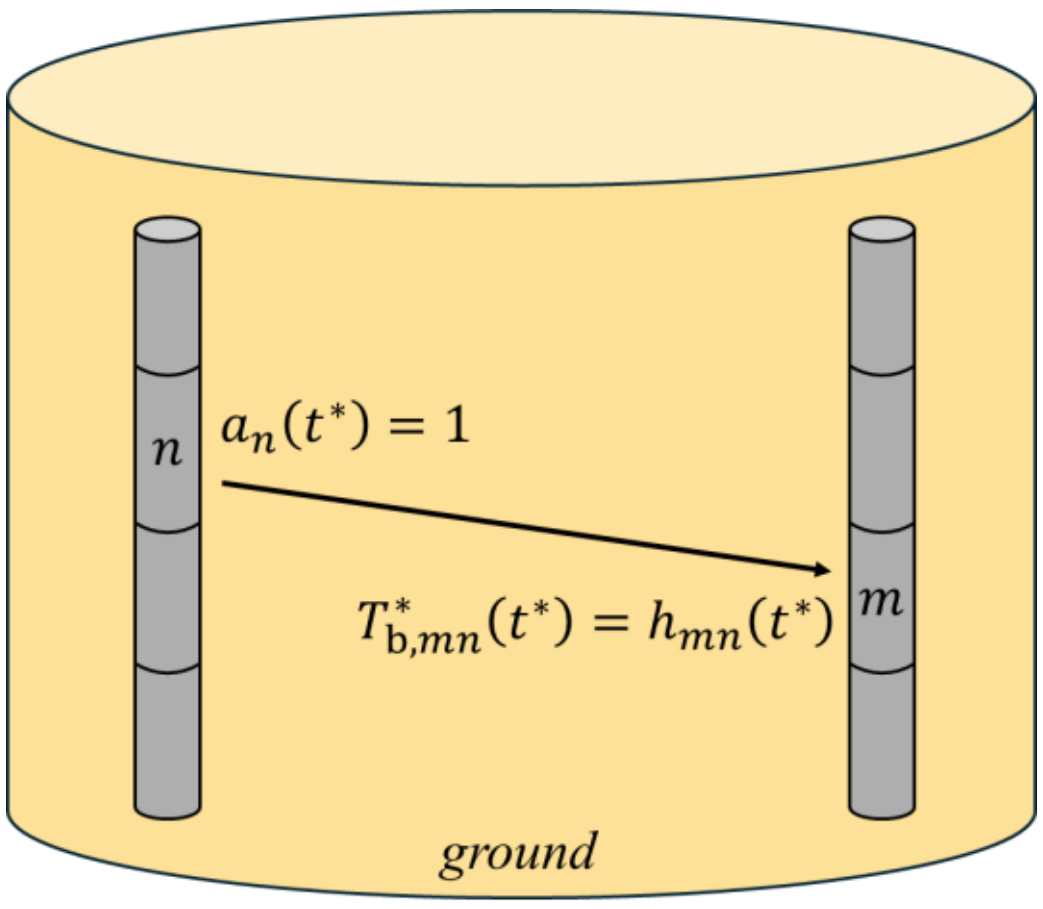

Figure 1: Effect of the thermal interaction between two segments of a bore field. If the $n$-th segment is subjected to a dimensionless heat load $a_n(t^*) = 1$, the contribution $T^*_{\mathrm{b},mn}(t^*)$ to the temperature increase of the $m$-th segment due to the $n$-th segment is $h_{mn}(t^*)$.

The functions $h_{mn}(t^*)$ can be called segment-to-segment FLS solutions. Their analytical expressions have been determined by Cimmino and Bernier [7]. We report them here below, in terms of the dimensionless quantities defined above:

$$h_{mn}(t^*) = \frac{1}{4\pi H^*}\int_{\frac{3}{\sqrt{4t^*}}}^{\infty} \frac{\exp\left(-{d^*_{mn}}^2 u^2\right)}{u^2}\{\mathrm{Ierf}[(D^*_m - D^*_n + H^*)u] + \mathrm{Ierf}[(D^*_m - D^*_n - H^*)u] - 2\mathrm{Ierf}[(D^*_m - D^*_n)u] + 2\mathrm{Ierf}[(D^*_m + D^*_n + H^*)u] - \mathrm{Ierf}[(D^*_m + D^*_n)u] - \mathrm{Ierf}[(D^*_m + D^*_n + 2H^*)u]\}\mathrm{d}u. \tag{12}$$

In Eq. (12), $H^*$ is the dimensionless length of each segment, $d^*_{mn}$ is the horizontal dimensionless distance between segment $m$ and segment $n$, $D^*_m$ and $D^*_n$ are, respectively, the dimensionless buried depths of the segments $m$ and $n$, and Ierf is the integral error function:

$$\mathrm{Ierf}(x) = x\,\mathrm{erf}(x) - \frac{1-e^{-x^2}}{\sqrt{\pi}}, \tag{13}$$

where

$$\mathrm{erf}(x) = \frac{2}{\sqrt{\pi}}\int_0^x \mathrm{e}^{-u^2}\,\mathrm{d}u. \tag{14}$$

If $m$ and $n$ belong to the same borehole, $d^*_{mn}$ is equal to the dimensionless radius of the borehole. For $t^* = 0$, one has $h_{mn}(0) = 0$. The values of $h_{mn}$ given by Eq. (12) correspond to those defined by Cimmino and Bernier divided by $2\pi$.

If the dimensionless heat load $a_n(t^*)$ of the segment $n$, instead, varies over time, in order to establish the contribution $T^*_{\mathrm{b},mn}(t^*)$ one can approximate $a_n(t^*)$ with a piecewise constant function. Assuming that one wants to study the bore field from the start of the heating, $t^*_0 = 0$, to a final instant $t^*_K$, one can choose a set of intermediate instants $t^*_0 < t^*_1 < t^*_2 < \cdots < t^*_K$ and denote by $a_n(t^*_i)$ the value of $a_n(t^*)$ in the time interval $t^*_i \le t^* < t^*_{i+1}$. This way, $a_n(t^*)$ can be expressed as

$$a_n(t^*) = \sum_{i=0}^{K-1}[U(t^* - t_i^*) - U(t^* - t_{i+1}^*)]a_n(t_i^*), \tag{15}$$

where $U$ is the Heaviside unit-step function:

$$U(x) = \begin{cases} 1 & \text{for} \quad x \geq 0 \\ 0 & \text{for} \quad x < 0 \end{cases}. \tag{16}$$

By applying the superposition of the effects in time, one finds that the contribution to the temperature increase of the $m$-th segment due to the heat load of the $n$-th segment, at any instant $t_k^*$, is given by

$$T_{\mathrm{b},mn}^*(t_k^*) = \sum_{i=0}^{k-1}[h_{mn}(t_k^* - t_i^*) - h_{mn}(t_k^* - t_{i+1}^*)]a_n(t_i^*). \tag{17}$$

Then, by adding the contributions due to all segments (see Eq. (11)), one obtains the dimensionless temperature of the $m$-th segment:

$$T_{\mathrm{b},m}^*(t_k^*) = \sum_{i=0}^{k-1}\sum_{n=0}^{N-1}[h_{mn}(t_k^* - t_i^*) - h_{mn}(t_k^* - t_{i+1}^*)]\, a_n(t_i^*). \tag{18}$$

Let us denote by $R_{\mathrm{b3D}}(t_k^*)$ the 3D borehole thermal resistance at the instant $t_k^*$, that can be assumed to be the same for all the BHEs. The temperature $T_{\mathrm{f},m}(t_k^*)$ of the fluid inside the $m$-th segment can be obtained from Eq. (4):

$$T_{\mathrm{f},m}(t_k^*) = T_{\mathrm{b},m}(t_k^*) + R_{\mathrm{b3D}}(t_k^*)a_m(t_{k-1}^*)q_{l0}, \tag{19}$$

i.e., in dimensionless form,

$$T_{\mathrm{f},m}^*(t_k^*) = T_{\mathrm{b},m}^*(t_k^*) + R_{\mathrm{b3D}}^*(t_k^*)a_m(t_{k-1}^*), \tag{20}$$

where $R_{\mathrm{b3D}}^* = k_{\mathrm{g}}R_{\mathrm{b3D}}$ is the dimensionless 3D borehole thermal resistance per unit length.

Eqs. (18) and (20) yield

$$T_{\mathrm{f},m}^*(t_k^*) = \sum_{i=0}^{k-1}\sum_{n=0}^{N-1} a_n(t_i^*)[h_{mn}(t_k^* - t_i^*) - h_{mn}(t_k^* - t_{i+1}^*)] + R_{\mathrm{b3D}}^*(t_k^*)a_m(t_{k-1}^*). \tag{21}$$

In order to have an isothermal fluid in the bore field, all the dimensionless temperatures $T_{\mathrm{f},m}^*(t_k^*)$ must be equal to $T_{\mathrm{f},0}^*(t_k^*)$, so that, for every segment $m$, the following condition must be fulfilled:

$$\sum_{i=0}^{k-1}\sum_{n=0}^{N-1} a_n(t_i^*)[h_{mn}(t_k^* - t_i^*) - h_{mn}(t_k^* - t_{i+1}^*)] + R_{\mathrm{b3D}}^*(t_k^*)a_m(t_{k-1}^*) = \sum_{i=0}^{k-1}\sum_{n=0}^{N-1} a_n(t_i^*)[h_{0n}(t_k^* - t_i^*) - h_{0n}(t_k^* - t_{i+1}^*)] + R_{\mathrm{b3D}}^*(t_k^*)a_0(t_{k-1}^*). \tag{22}$$

Moreover, since the mean heat load per unit length is equal to $\dot{q}_{l0}$, the dimensionless mean heat load per unit length is equal to 1. Thus, for every value of $k$, one has

$$\sum_{n=0}^{N-1} a_n(t_{k-1}^*) = N. \tag{23}$$

Eqs. (22) and (23) can be merged into a single equation by introducing the Kronecker delta function

$$\delta_{mn} = \begin{cases} 1 & \text{for} \quad m = n \\ 0 & \text{for} \quad m \neq n \end{cases}. \tag{24}$$

Note that Eq. (22) is an identity for $m = 0$. Therefore, if Eq (23) is multiplied for $\delta_{m0}$ and added to Eq. (22), one obtains a single equation that coincides with Eq. (22) for $m \neq 0$ and with Eq. (23) for $m = 0$:

$$\sum_{i=0}^{k-1}\sum_{n=0}^{N-1} a_n(t_i^*)[h_{mn}(t_k^* - t_i^*) - h_{mn}(t_k^* - t_{i+1}^*)] + R_{\mathrm{b3D}}^*(t_k^*)a_m(t_{k-1}^*) + \delta_{m0}\sum_{n=0}^{N-1} a_n(t_{k-1}^*) = \sum_{i=0}^{k-1}\sum_{n=0}^{N-1} a_n(t_i^*)[h_{0n}(t_k^* - t_i^*) - h_{0n}(t_k^* - t_{i+1}^*)] + R_{\mathrm{b3D}}^*(t_k^*)a_0(t_{k-1}^*) + \delta_{m0}N. \tag{25}$$

Eq. (25) can be solved by subsequent iterations. For $k = 1$, Eq. (25) yields a system of $N$ linear equations in the $N$ unknowns $a_n(t_0^*)$. Once the system is solved for $k = 1$, the obtained values of $a_n(t_0^*)$ can be substituted back into Eqs. (25), allowing to solve the system for $k = 2$. This yields the values of $a_n(t_1^*)$. The process can then be repeated until the values of $a_n(t_{k-1}^*)$ are obtained.

To ease the solution, Eq. (25) can be written in matrix form by extracting from the sum in $i$ the term obtained for $i = k - 1$, and by exploiting the mathematical properties of the Kronecker delta function. The end result is an equation of the form

$$\mathrm{A}_{mn} a_n(t_{k-1}^*) = b_m, \tag{26}$$

where $\mathrm{A}_{mn}$ is the $N \times N$ matrix

$$\mathrm{A}_{mn} = h_{mn}(t_k^* - t_{k-1}^*) - h_{0n}(t_k^* - t_{k-1}^*) + R_{\mathrm{b3D}}^*(t_k^*)(\delta_{mn} - \delta_{0n}) + \delta_{m0}, \tag{27}$$

and $b_m$ is the vector

$$b_m = \delta_{m0} N - \sum_{i=0}^{k-2} \sum_{n=0}^{N-1} a_n(t_i^*)[h_{mn}(t_k^* - t_i^*) - h_{mn}(t_k^* - t_{i+1}^*) - h_{0n}(t_k^* - t_i^*) + h_{0n}(t_k^* - t_{i+1}^*)]. \tag{28}$$

Note that $b_m$ depends on the values $a_n(t_i^*)$ with $i$ ranging from 0 to $k - 2$, that are obtained in the preceding iterations. For $k = 1$, $b_m$ is simply equal to $\delta_{m0} N$.

Once the dimensionless heat loads $a_n(t_i^*)$ are obtained, they can be substituted into Eq. (21) to get dimensionless temperature of the fluid, $T_\mathrm{f}^*(t_k^*)$, which is the same for all segments, namely the *ftg-function*. Lastly, one can average Eq. (20) over all the segments and use it to obtain the mean dimensionless temperature of the external surface of the BHEs, $T_\mathrm{b}^*(t_k^*)$, i.e., the *g-function*. Since the mean dimensionless heat load is equal to 1, one simply obtains

$$T_\mathrm{b}^*(t_k^*) = T_\mathrm{f}^*(t_k^*) - R_{\mathrm{b3D}}^*(t_k^*). \tag{29}$$

## 3. Optimization of the computation time

The model proposed in section 2 can be used to write a computer program that calculates the dimensionless heat load of each borehole segment, the *ftg-function,* and the *g-function* of the bore field. Indeed, the segment-to-segment FLS solutions, $h_{mn}(t^*)$, can be obtained by numerically computing the integral appearing in Eq. (12), for instance by using the trapezoidal rule. Once the $h_{mn}(t^*)$ functions are known, the systems of equations given by Eqs. (26-28) can be solved iteratively for increasing values of $k$, for instance by using the Gaussian elimination method. There are, however, two main issues that may arise in the realization of such a program, which will be addressed in this section.

The first issue concerns the computation time of the $h_{mn}(t^*)$ functions. Let us assume, for instance, that the object of our study is a rectangular bore field, namely a 2D array of BHEs with a constant distance between the columns, $B_x$, and a constant distance between the rows, $B_y$ (see Fig. 2).

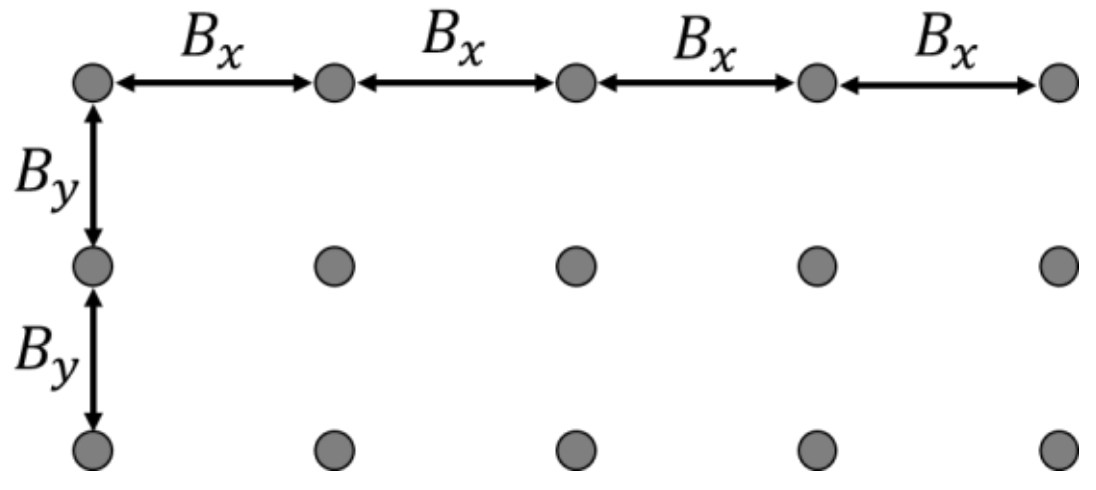

Figure 2: Cross section of a rectangular 3 × 5 bore field. The distance between the columns, $B_x$, and the distance between the rows, $B_y$, are both constant.

Assuming that each borehole is divided into 100 segments, the total number of segments in the example of Fig. 2 is $N = 1{,}500$. If the $h_{mn}(t^*)$ functions are computed for every value of $m$ and for every value of $n$, the number of integrals that the program needs to evaluate is $N^2 = 2{,}225{,}000$, that is a couple of million integrals. Fortunately, as shown by Cimmino in Ref. [9], this issue can be solved rather easily. Indeed, the $h_{mn}(t^*)$ functions in Eq. (12) can be written as a sum of two terms:

$$h_{mn}(t^*) = h^+_{mn}(t^*) + h^-_{mn}(t^*), \tag{30}$$

where

$$h^+_{mn}(t^*) = \frac{1}{4\pi H^*}\int_{\frac{3}{\sqrt{4t^*}}}^{\infty} \frac{\exp\left(-{d^*_{mn}}^2 u^2\right)}{u^2}\{2\mathrm{Ierf}[(D^*_m + D^*_n + H^*)u] - \mathrm{Ierf}[(D^*_m + D^*_n)u] - \mathrm{Ierf}[(D^*_m + D^*_n + 2H^*)u]\}\mathrm{d}u \tag{31}$$

and

$$h^-_{mn}(t^*) = \frac{1}{4\pi H^*}\int_{\frac{3}{\sqrt{4t^*}}}^{\infty} \frac{\exp\left(-{d^*_{mn}}^2 u^2\right)}{u^2}\{\mathrm{Ierf}[(D^*_m - D^*_n + H^*)u] + \mathrm{Ierf}[(D^*_m - D^*_n - H^*)u] - 2\mathrm{Ierf}[(D^*_m - D^*_n)u]\}\mathrm{d}u. \tag{32}$$

Note that $h^+_{mn}(t^*)$ depends exclusively on the horizontal distance $d^*_{mn}$ between the two segments and on the sum $D^*_m + D^*_n$ of the two depths. Similarly, $h^-_{mn}(t^*)$ depends exclusively on the horizontal distance $d^*_{mn}$ between the two segments and on the difference $D^*_m - D^*_n$ between the two depths.

So, instead of computing $h_{mn}(t^*)$ for every value of $m$ and $n$, it is sufficient to compute $h^+_{mn}(t^*)$ for every possible value of $d^*_{mn}$ and of the sum $D^*_m + D^*_n$, and $h^-_{mn}(t^*)$ for every possible value of $d^*_{mn}$ and of the difference $D^*_m - D^*_n$. In the example discussed above, for instance, there are only 15 possible values of the horizontal distance $d^*_{mn}$, 199 possible values of the sum $D^*_m + D^*_n$, and 199 possible values of the difference $D^*_m - D^*_n$. Thus, by exploiting this method, the number of integrals that the program needs to evaluate drops from $2{,}225{,}000$ to $15 \times 199 \times 2 = 5{,}970$.

The second issue concerns the computation time of the linear systems of equations. Indeed, the number of operations required to solve a $N \times N$ linear system with Gaussian elimination is proportional to $N^3$. This means that, for large bore fields, the time required to solve the linear systems increases significantly. Luckily, the computation time of the linear systems can be greatly reduced by exploiting the geometrical symmetries that are commonly present in bore fields.

All rectangular bore fields, for instance, have two symmetry planes. If two segments, located at the same depth, are positioned symmetrically with respect to a symmetry plane, we will state that the two segments are equivalent. This concept is illustrated in Fig. 3, where equivalent segments are depicted in the same color.

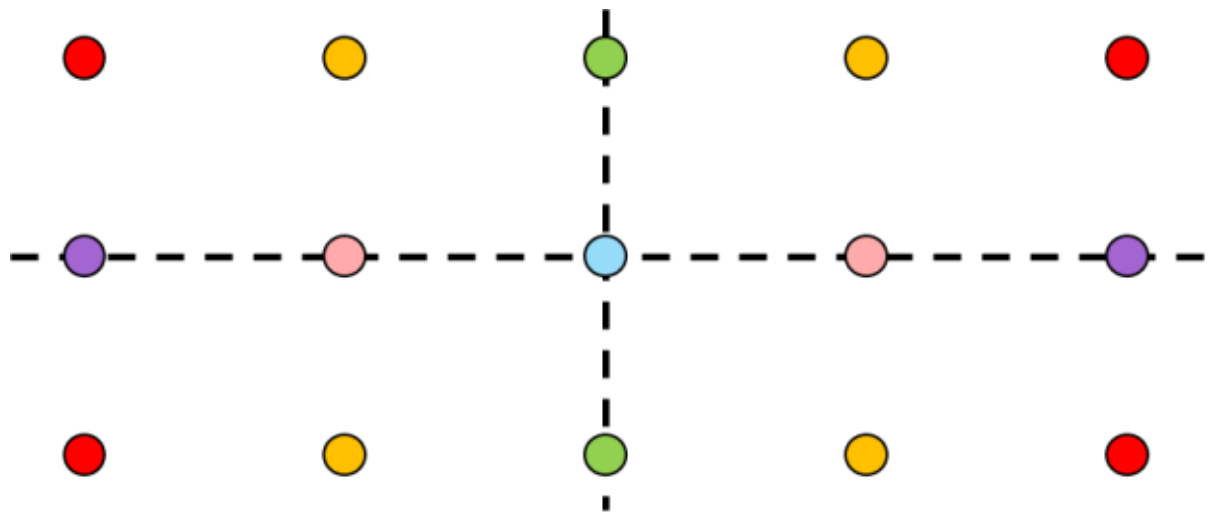

Figure 3: Cross section of a 3 × 5 rectangular bore field. The two dashed lines represent the symmetry planes of the field. Equivalent segments are depicted in the same color.

If a bore field has one or more symmetry planes, borehole segments can be grouped into sets of equivalent segments. In the example shown in Fig. 3, assuming that each borehole is divided in 100 segments, we can group the 1500 total segments in 200 sets of four segments, 300 sets of two segments, and 100 sets containing just one segment, since the segments of the central borehole in Fig. 3 do not have any equivalent segment. We can then denote as $P$ the total number of sets (in our example, $P = 600$), and we can index each set with an integer $p$, ranging from 0 to $P - 1$. This arrangement of the segments into sets is particularly useful because, due to the symmetries of the bore field, equivalent segments must have the same physical attributes.

As we explained in section 2, the mean dimensionless temperature $T^*_{\mathrm{f},m}$ of the fluid inside a segment $m$ is given by Eq. (21). If the borehole segments are divided into sets, the sum for $n = 0$ to $N - 1$ appearing in Eq. (21) can be split into a sum for $p = 0$ to $P - 1$ and a sum for $n \in p$:

$$T^*_{\mathrm{f},m}(t^*_k) = \sum_{i=0}^{k-1}\sum_{p=0}^{P-1}\sum_{n\in p} a_n(t^*_i)[h_{mn}(t^*_k - t^*_i) - h_{mn}(t^*_k - t^*_{i+1})] + R^*_{\mathrm{b3D}}(t^*_k)a_m(t^*_{k-1}). \quad (33)$$

Since the segments $n$ belonging to the same set $p$ are equivalent, the heat loads $a_n(t^*_i)$ for $n \in p$ must all have the same value, which we will denote as $a_p(t^*_i)$. By setting this condition in Eq. (33), we get

$$T^*_{\mathrm{f},m}(t^*_k) = \sum_{i=0}^{k-1}\sum_{p=0}^{P-1} a_p(t^*_i)\sum_{n\in p}[h_{mn}(t^*_k - t^*_i) - h_{mn}(t^*_k - t^*_{i+1})] + R^*_{\mathrm{b3D}}(t^*_k)a_m(t^*_{k-1}). \quad (34)$$

Now, in order to have isothermal fluid in the bore field, we should require that the fluid temperature $T^*_{\mathrm{f},m}(t^*_k)$ is the same for all segments. However, we already set into Eq. (34) the condition that equivalent segments have the same heat load, which entails that equivalent segments also have the same fluid temperature. Therefore, it is sufficient to require that non-equivalent segments have the same fluid temperature. We can then create a collection of segments, $w$, which contains only the first segment $m$ of each set $p$ (for a total of $P$ segments), and require that $T^*_{f,m}(t^*_k) = T^*_{f,0}(t^*_k)$ for every segment $m$ inside $w$. By doing so, we get

$$\sum_{i=0}^{k-1}\sum_{p=0}^{P-1} a_p(t^*_i)\sum_{n\in p}[h_{mn}(t^*_k - t^*_i) - h_{mn}(t^*_k - t^*_{i+1})] + R^*_{b3D}(t^*_k)a_m(t^*_{k-1}) = \sum_{i=0}^{k-1}\sum_{p=0}^{P-1} a_p(t^*_i)\sum_{n\in p}[h_{0n}(t^*_k - t^*_i) - h_{0n}(t^*_k - t^*_{i+1})] + R^*_{b3D}(t^*_k)a_0(t^*_{k-1}) \quad (35)$$

for every $m \in w$.

Moreover, since the mean heat load per unit length is equal to $\dot{q}_{l0}$, the dimensionless mean heat load per unit length is equal to 1. Thus, for every value of $k$, one has

$$\sum_{p=0}^{P-1}\sum_{n\in p} a_n(t^*_{k-1}) = \sum_{p=0}^{P-1} a_p(t^*_{k-1})\sum_{n\in p} 1 = N. \quad (36)$$

Eqs. (35) and (36) can be merged into a single equation, in the same way as we did in section 2:

$$\sum_{i=0}^{k-1}\sum_{p=0}^{P-1} a_p(t^*_i)\sum_{n\in p}[h_{mn}(t^*_k - t^*_i) - h_{mn}(t^*_k - t^*_{i+1})] + R^*_{b3D}(t^*_k)a_m(t^*_{k-1}) + \delta_{m0}\sum_{p=0}^{P-1} a_p(t^*_{k-1})\sum_{n\in p} 1 = \sum_{i=0}^{k-1}\sum_{p=0}^{P-1} a_p(t^*_i)\sum_{n\in p}[h_{0n}(t^*_k - t^*_i) - h_{0n}(t^*_k - t^*_{i+1})] + R^*_{b3D}(t^*_k)a_0(t^*_{k-1}) + \delta_{m0}N \quad (37)$$

Similarly to Eq. (25), Eq. (37) can be solved by subsequent iterations. For every value of $k$, Eq. (37) yields a system of $P$ linear equations in the $P$ unknowns $a_p(t^*_{k-1})$. Like Eq. (25), Eq. (37) can also be written in matrix form. The end result is an equation of the form

$$\mathrm{A}_{mp}a_p(t^*_{k-1}) = b_m, \quad (38)$$

where $\mathrm{A}_{mp}$ (with $m \in w$) is the $P \times P$ matrix

$$\mathrm{A}_{mp} = \sum_{n\in p}[h_{mn}(t^*_k - t^*_{k-1}) - h_{0n}(t^*_k - t^*_{k-1}) + R^*_{b3D}(t^*_k)(\delta_{mn} - \delta_{0n}) + \delta_{m0}], \quad (39)$$

and $b_m$ (with $m \in w$) is the vector

$$b_m = \delta_{m0}N - \sum_{i=0}^{k-2}\sum_{p=0}^{P-1} a_p(t^*_i)\sum_{n\in p}[h_{mn}(t^*_k - t^*_i) - h_{mn}(t^*_k - t^*_{i+1}) - h_{0n}(t^*_k - t^*_i) + h_{0n}(t^*_k - t^*_{i+1})]. \quad (40)$$

Note that $b_m$ depends on the values $a_p(t^*_i)$ obtained in the preceding iterations, with the convention that, for $k = 1$, $b_m$ is simply equal to $\delta_{m0}N$.

Once the dimensionless heat loads $a_p(t^*_i)$ are obtained, they can be substituted into Eq. (34) to get dimensionless temperature of the fluid, $T^*_f(t^*_k)$, while the mean dimensionless temperature of the external surface of the BHEs, $T^*_b(t^*_k)$, can be obtained from Eq. (29).

The computational advantage of using Eqs. (38-40) instead of Eqs. (26-28) can be very significant. As we already mentioned, if the Gaussian elimination method is used, the number of operations required to solve a linear system of equations is proportional to the number of equations to the third power. Thus, if Eqs. (38-40) are used instead of Eqs. (26-28), the number of operations required to solve the linear systems is reduced by a factor $f = (N/P)^3$. In the example shown in Fig. 3, $N/P = 1{,}500/600 = 2.5$, so $f \approx 15.6$. For rectangular bore fields with an even number of rows and an even number of columns, the advantage is even greater, with a factor $f = 64$.

Moreover, the method proposed in this section for the solution of the linear systems does not only work for rectangular bore fields. It applies, more generally, to bore fields of any shape, provided that they have at least one symmetry plane: of course, the greater the number of symmetry planes, the greater the advantage. Some examples of bore fields of various shapes are shown in Fig. 4, where the factor $f$ is also indicated.

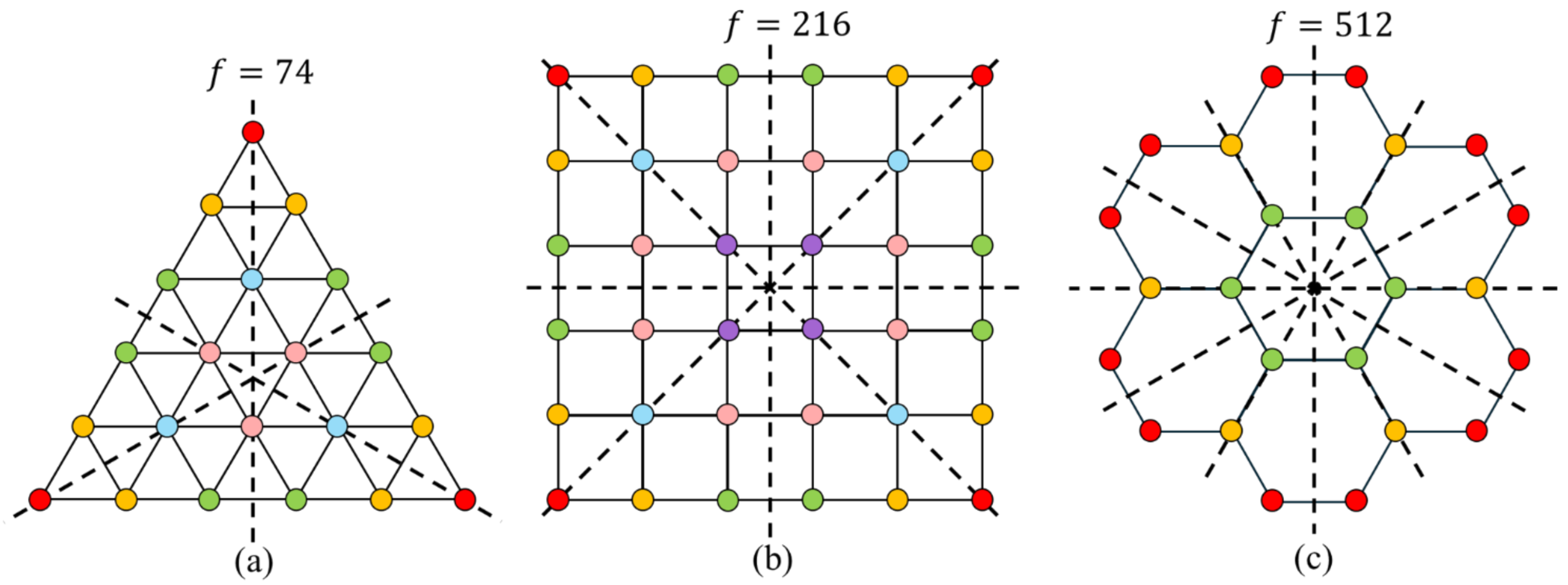

Figure 4: Cross section of bore fields of various shapes: equilateral triangular (a), square (b) and regular hexagonal (c). By exploiting the symmetry planes (dashed lines), the number of operations required to solve the linear systems is reduced by a factor $f$.

**4. Description of the simulation program**

A simulation program was written in C++ implementing the optimization methods described in section 3, as well as some common programming techniques such as multiprocessing. The program can simulate bore fields of any shape, provided that the boreholes are fed in parallel with an equal inlet fluid temperature, have the same length $H_\mathrm{b}$, and are buried at the same depth $D$. The *g-function*, the *ftg-function* and the dimensionless heat loads of the segments are evaluated at evenly distributed instants on a logarithmic time scale, namely for equally spaced values of $\ln(t^*)$.

As for the 3D borehole thermal resistance, the program has two different options: it can either estimate a constant value of $R_\mathrm{b3D}$ based on the characteristics of the BHEs and on the properties of the fluid, or it can read a vector of values $R_\mathrm{b3D}(t_i^*)$ from an input file supplied by the user.

In the first option, the value of $R_\mathrm{b3D}$ in quasi-stationary conditions is estimated as the arithmetic average of $R_\mathrm{b}$ and of the effective borehole thermal resistance, $R_\mathrm{beff}$, from the following input data: properties of the fluid, namely density, $\rho_\mathrm{f}$, specific heat capacity at constant pressure, $c_{p\mathrm{f}}$, thermal conductivity, $k_\mathrm{f}$, and dynamic viscosity, $\mu_\mathrm{f}$; geometry of the BHEs, namely BHE radius, $r_\mathrm{b}$, half shank spacing, $s$ (i.e., distance between pipe axis and BHE axis), internal radius, $r_\mathrm{i}$, and external radius, $r_\mathrm{e}$, of the pipes; and physical properties of the bore field, namely thermal conductivity of the ground, $k_\mathrm{g}$, of the grout, $k_\mathrm{gt}$, and of the pipe, $k_\mathrm{p}$, and volume flow rate in each BHE, $\dot{V}$.

For single U-tube BHEs, $R_{\mathrm{b}}$ and $R_{\mathrm{eff}}$ are determined with the method proposed by Hellström [35]. The convection coefficient, $h$, is determined by the Churchill correlation [36], with uniform heat flux. The thermal resistance of the pipe per unit length is given by

$$R_{\mathrm{p}} = \frac{1}{2\pi r_{\mathrm{i}} h} + \frac{1}{2\pi k_{\mathrm{p}}} \ln\left(\frac{r_{\mathrm{e}}}{r_{\mathrm{i}}}\right). \tag{41}$$

The 2D borehole thermal resistance, $R_{\mathrm{b}}$, and the total internal thermal resistance between the pipes, $R_{\mathrm{a}}$, are determined by Hellström's expressions in the line-source approximation,

$$R_{\mathrm{b}} = \frac{1}{4\pi k_{\mathrm{gt}}} \left[\ln\left(\frac{r_{\mathrm{b}}}{r_{\mathrm{e}}}\right) + \ln\left(\frac{r_{\mathrm{b}}}{2s}\right) + \sigma \ln\left(\frac{r_{\mathrm{b}}^4}{r_{\mathrm{b}}^4 - s^4}\right)\right] + \frac{R_{\mathrm{p}}}{2}, \tag{42}$$

$$R_{\mathrm{a}} = \frac{1}{\pi k_{\mathrm{gt}}} \left[\ln\left(\frac{2s}{r_{\mathrm{e}}}\right) + \sigma \ln\left(\frac{r_{\mathrm{b}}^2 + s^2}{r_{\mathrm{b}}^2 - s^2}\right)\right] + 2R_{\mathrm{p}}, \tag{43}$$

where $\sigma$ is a dimensionless parameter, defined as

$$\sigma = \frac{k_{\mathrm{gt}} - k_{\mathrm{g}}}{k_{\mathrm{gt}} + k_{\mathrm{g}}}. \tag{44}$$

The effective borehole thermal resistance is given by

$$R_{\mathrm{beff}} = \eta \coth(\eta)\, R_{\mathrm{b}}, \tag{45}$$

where $\eta$ is a dimensionless parameter that can be expressed as in Ref. [37]:

$$\eta = \frac{H_{\mathrm{b}}}{c_{p\mathrm{f}}\, \rho_{\mathrm{f}}\, \dot{V}} \frac{1}{\sqrt{R_{\mathrm{a}} R_{\mathrm{b}}}}\,. \tag{46}$$

For double U-tube BHEs, $R_{\mathrm{b}}$ and $R_{\mathrm{beff}}$ are determined by the method proposed by Zeng et al. [38] and employed by Zanchini and Jahanbin [39] for the validation of finite-element simulations. Due to misprints in Eq. (10) of Ref. [38] and in Eqs. (24) and (25) of Ref. [39], all the necessary equations are reported here.

The pipes are numbered from 1 to 4, as illustrated in Fig. 2 of Ref. [38]. Pipe 2 is close to pipe 1, and pipe 3 is opposite to pipe 1. The inlet is in pipes 1 and 2, in parallel. In the line source approximation, the thermal resistances between pipe 1 and the outer surface of the borehole, between pipe 1 and pipe 2, and between pipe 1 and pipe 3 are, respectively,

$$R_{11} = \frac{1}{2\pi k_{\mathrm{gt}}} \left[\ln\left(\frac{r_{\mathrm{b}}}{r_{\mathrm{e}}}\right) - \sigma \ln\left(\frac{r_{\mathrm{b}}^2 - s^2}{r_{\mathrm{b}}^2}\right)\right] + R_{\mathrm{p}}, \tag{47}$$

$$R_{12} = \frac{1}{2\pi k_{\mathrm{gt}}} \left[\ln\left(\frac{r_{\mathrm{b}}}{\sqrt{2}\, s}\right) - \frac{\sigma}{2} \ln\left(\frac{r_{\mathrm{b}}^4 + s^4}{r_{\mathrm{b}}^4}\right)\right], \tag{48}$$

$$R_{13} = \frac{1}{2\pi k_{\mathrm{gt}}} \left[\ln\left(\frac{r_{\mathrm{b}}}{2\, s}\right) - \sigma \ln\left(\frac{r_{\mathrm{b}}^2 + s^2}{r_{\mathrm{b}}^2}\right)\right]. \tag{49}$$

The 2D borehole thermal resistance is given by

$$R_{\mathrm{b}} = \frac{R_{11} + R_{13} + 2R_{12}}{4}. \tag{50}$$

The dimensionless bulk fluid temperature is defined as

$$\Theta(z^*) = \frac{T(z^*) - T_{\mathrm{b}}}{T_{\mathrm{in}} - T_{\mathrm{b}}}, \tag{51}$$

where $T_{\mathrm{in}}$ is the inlet fluid temperature, $T_{\mathrm{b}}$ is the temperature of the external surface of the borehole (assumed to be uniform), and $z^* = z/H_{\mathrm{b}}$ is the dimensionless distance from the BHE top. The distribution of the dimensionless bulk fluid temperature in the downward flow, $\Theta_{\mathrm{d}}$, and in the upward flow, $\Theta_{\mathrm{u}}$, can be expressed as

$$\Theta_{\mathrm{d}}(z^*) = \frac{C\cosh[S(1-z^*)]+\sinh[S(1-z^*)]}{C\cosh(S)+\sinh(S)}, \tag{52}$$

$$\Theta_{\mathrm{u}}(z^*) = \frac{C\cosh[S(1-z^*)]-\sinh[S(1-z^*)]}{C\cosh(S)+\sinh(S)}, \tag{53}$$

where $C$ and $S$ are dimensionless parameters defined as follows:

$$C = \sqrt{2\frac{R_{12}+R_{13}}{R_{11}-R_{13}}+1}, \tag{54}$$

$$S = \frac{H_{\mathrm{b}}}{2\,c_{p\mathrm{f}}\,\rho_{\mathrm{f}}\,\dot{V}\,R_{\mathrm{b}}}C. \tag{55}$$

The effective borehole thermal resistance is given by

$$R_{\mathrm{beff}} = \frac{H_{\mathrm{b}}}{2\,c_{p\mathrm{f}}\,\rho_{\mathrm{f}}\,\dot{V}}\,\frac{1+\Theta_{\mathrm{u}}(0)}{1-\Theta_{\mathrm{u}}(0)} = S\coth(S)\,R_{\mathrm{b}}. \tag{56}$$

In the second option, the user needs to supply to the program a vector of values of $R_{\mathrm{b3D}}(t_i^*)$, which can be obtained through a suitable numerical simulation of a single BHE. To determine the errors in the approximate analytical estimates of the quasi-stationary value of $R_{\mathrm{b3D}}$, finite-element simulations of 4 single U-tube BHEs and 4 double U-tube BHEs with length 100m and different cross sections have been performed.

The simulations were performed with COMSOL Multiphysics, and the energy balance along the pipes was simulated by the Pipe Flow Module. The pipe walls were sketched as solid cylinders with a reduced volumetric heat capacity, as recommended in Ref. [17]. Other details of the simulations are the same as those explained in Ref. [17], with the kind of meshes denoted there as Mesh 2.

In these simulations, the polyethylene pipes of the single U-tube BHEs have external radius 20 mm and internal radius 16.3 mm; those of the double U-tube BHEs have external radius 16 mm and internal radius 13 mm; the thermal conductivity and the volumetric heat capacity of polyethylene are, respectively, 0.4 W/(m K) and 1.824 MJ/(m$^3$K). The single U-tube BHEs have shank spacing equal to 94 mm and 54 mm; the double U-tube BHEs have shank spacing equal to 102 mm and 85 mm. For each geometry, two thermal conductivities of the grout were considered, namely 1.6 W/(m K) and 1.0 W/(m K), and a volumetric heat capacity of the grout equal to that of the ground. The convection coefficient was determined by the Churchill correlation [36].

In these and in all the other simulations presented later in this paper, the BHEs considered have radius 76 mm and are buried at a depth of 1.8 m; the ground has thermal conductivity 1.8 W/(m K) and volumetric heat capacity 3.0 MJ/(m$^3$K), that correspond to a thermal diffusivity $\alpha_{\mathrm{g}} =$ $0.6 \times 10^{-6}$ m$^2$/s; the operating fluid is water with volume flow rate 14 liters per minute and properties evaluated at 20 °C, with values taken from Ref. [40].

The asymptotic values of $R_{b3D}$ determined by the finite-element simulations have been compared with the stationary values of $R_{b3D}$ estimated by the analytical method. The results of the comparison are illustrated in Table 1. The relative error in the approximate estimate of $R_{b3D}$ is between -1.168% and +0.823% in all cases, except for the BHE 2U102 1.6, where it is +3.075%.

Table 1: Comparison between the asymptotic values of $R_{b3D}$ determined by finite-element simulations and the constant values estimated analytically.

| Kind of BHE | BHE symbol | $R_{b3D}$ numerical | $R_{b3D}$ analytical | % error |
|---|---|---|---|---|
| Single U-tube, $2s = 94$ mm, $k_{gt} = 1.6$ W/(m K) | U94 1.6 | 0.1024 | 0.1030 | 0.608 |
| Single U-tube, $2s = 94$ mm, $k_{gt} = 1.0$ W/(m K) | U94 1.0 | 0.1334 | 0.1327 | -0.553 |
| Single U-tube, $2s = 54$ mm, $k_{gt} = 1.6$ W/(m K) | U54 1.6 | 0.1338 | 0.1320 | -1.324 |
| Single U-tube, $2s = 54$ mm, $k_{gt} = 1.0$ W/(m K) | U54 1.0 | 0.1839 | 0.1808 | -1.690 |
| Double U-tube, $2s = 102$ mm, $k_{gt} = 1.6$ W/(m K) | 2U102 1.6 | 0.0613 | 0.0632 | 3.075 |
| Double U-tube, $2s = 102$ mm, $k_{gt} = 1.0$ W/(m K) | 2U102 1.0 | 0.0818 | 0.0820 | 0.261 |
| Double U-tube, $2s = 85$ mm, $k_{gt} = 1.6$ W/(m K) | 2U85 1.6 | 0.0767 | 0.0773 | 0.823 |
| Double U-tube, $2s = 85$ mm, $k_{gt} = 1.0$ W/(m K) | 2U85 1.0 | 0.1056 | 0.1044 | -1.168 |

The time evolution of $R_{b3D}$ for the Single U-tube BHEs, in the range $-20 \le \ln(t^*) \le 0$ is illustrated in Fig. 5. The figure shows that the asymptotic values are nearly reached at $\ln(t^*) = -10$. The values reported in Table 1 are taken at $\ln(t^*) = -3$.

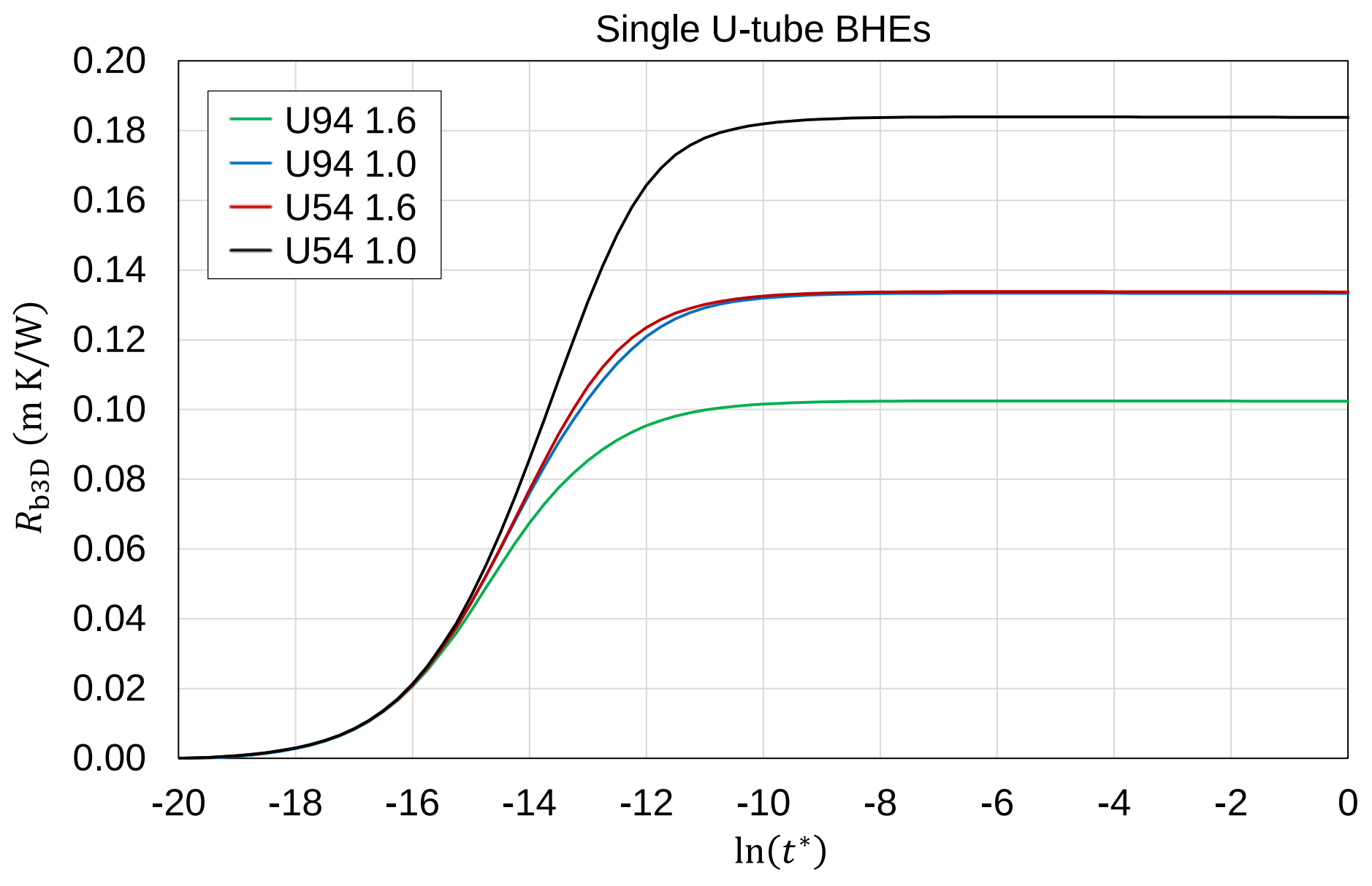

Figure 5: Time evolution of $R_{b3D}$ for the single U-tube BHEs U94 1.6, U94 1.0, U54 1.6, U54 1.0, in the time range $-20 \le \ln(t^*) \le 0$.

## 5. Effect of the number of segments on the *g-function*

The effect of the number of segments on the *g-function* has been analyzed by calculating the *g-functions* of a $4 \times 4$ bore field and of an $8 \times 8$ bore field with different numbers of segments. The single U-tube BHE U94 1.6 was chosen for both bore fields, with BHEs of length $H_\mathrm{b} = 100$ m, buried at a depth $D = 1.8$ m, and spaced at a distance $B = 7.5$ m from each other, both on the $x$ and the $y$ axis. The thermal conductivity of the ground was set to 1.8 $W/(\mathrm{m\,K})$. A constant value of the 3D borehole thermal resistance was estimated by the program, equal to 1.030 m K/W. The simulations have been performed with 12, 25, 50, 100 and 200 segments, for 89 time instants evenly distributed in the range $-16 \leq \ln(t^*) \leq 6$, namely with a time step $\Delta\ln(t^*) = 0.25$. With the considered values of the BHE length, $H_\mathrm{b} = 100$ m, and of the thermal diffusivity of the ground, $\alpha_\mathrm{g} = 0.6 \times 10^{-6}$ $\mathrm{m^2/s}$, this range corresponds to the time interval from 3.47 minutes to 23,690 years.

The *g-functions* obtained for the $4 \times 4$ bore field are illustrated in Fig. 6. The figure shows that, while the calculation with 12 segments yields an appreciable overestimation of the *g-function* in the long term, the *g-functions* obtained with 25, 50, 100 and 200 segments are very close to each other. The percent difference between the *g-functions* calculated with lower numbers of segments and the *g-function* calculated with 200 segments is also illustrated in Fig. 6, with values on the right axis. The figure shows that the percent difference is an increasing function of time. The percent difference between the *g-function* computed with 100 segments and that computed with 200 segments is the lowest, with a maximum deviation of 0.05%.

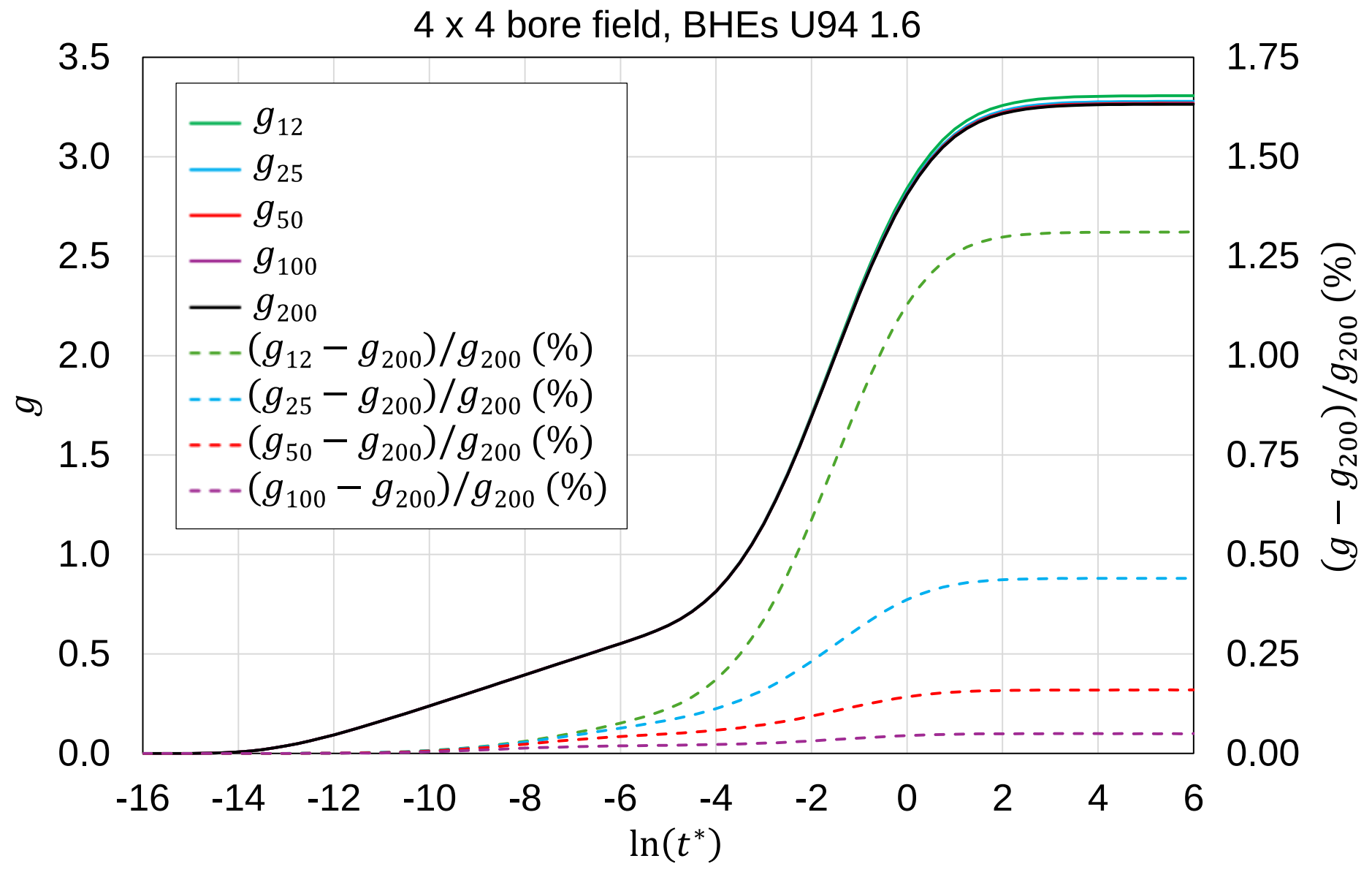

Figure 6: *g-functions* of a $4 \times 4$ bore field computed with 12, 25, 50, 100 and 200 segments, and percent differences between the first four curves and the last one. The simulations were performed with BHEs U94 1.6, $H_\mathrm{b} = 100$ m, $D = 1.8$ m, $B = 7.5$ m, $k_\mathrm{g} = 1.8$ W/(m K).

Analogous plots are shown for the 8 × 8 bore field in Fig. 7. For this bore field, the percent difference between the *g-function* computed with 100 segments and that computed with 200 segments is still rather low, with a maximum deviation of 0.08%. For this reason, the number of segments was fixed to 100 for all the subsequent simulations.

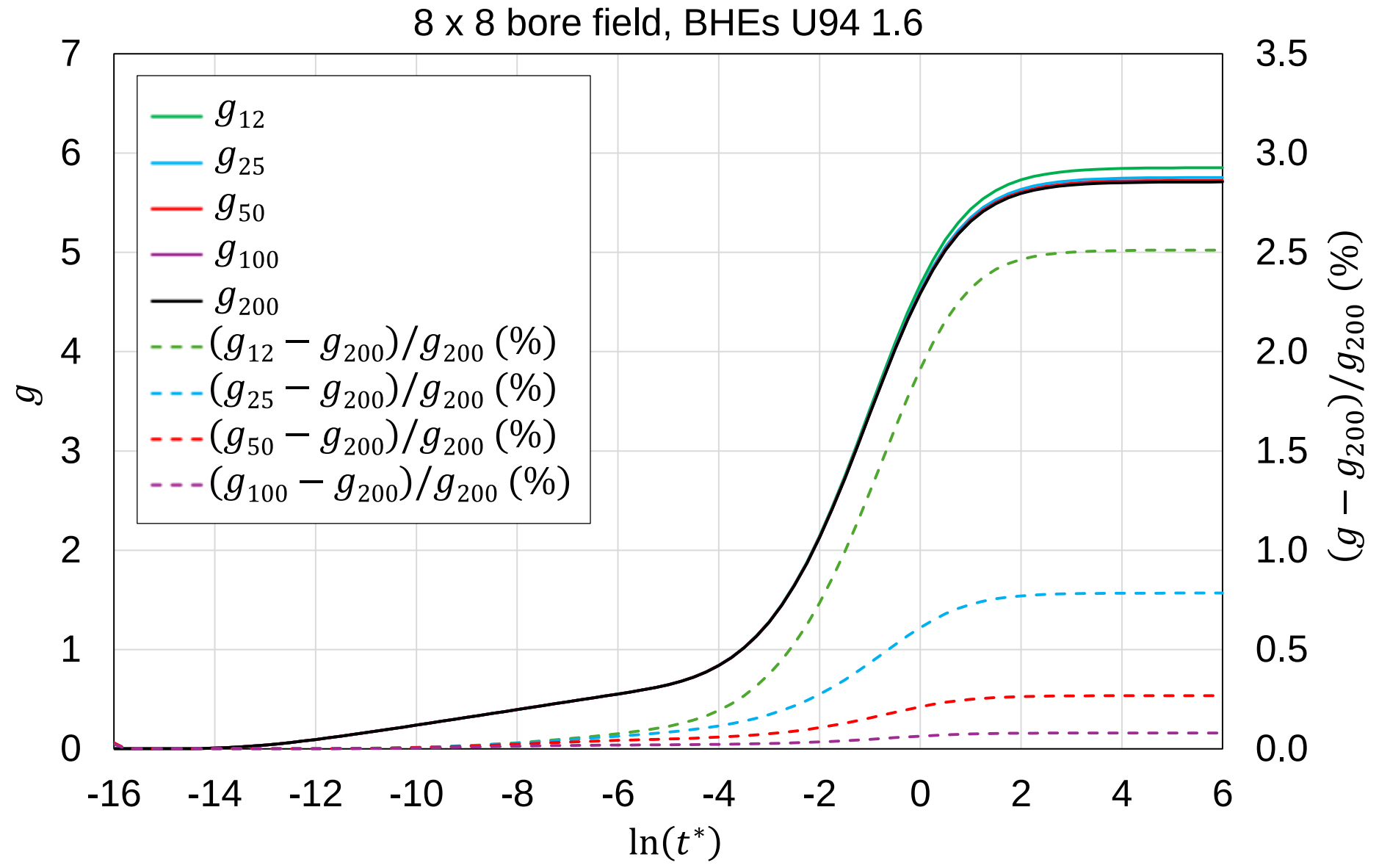

Figure 7: *g-functions* of a 8 × 8 bore field computed with 12, 25, 50, 100 and 200 segments, and percent differences of the first four curves with respect to the last one. The simulations were performed with BHEs U94 1.6, $H_b = 100$ m, $D = 1.8$ m, $B = 7.5$ m, $k_g = 1.8$ W/(m K).

The program was run on a Linux PC with processor Intel core i9-14900HX and 64 GB of RAM. The total runtime of each simulation is reported in Table 2. Thanks to the optimization methods presented in section 3, the execution times are all very short.

Table 2: simulation runtime for the 4 × 4 bore field and the 8 × 8 bore field.

| Number of segments | 4 × 4 bore field - runtime [s] | 8 × 8 bore field - runtime [s] |
|---|---|---|
| 12 | 0.46 | 1.59 |
| 25 | 0.94 | 4.36 |
| 50 | 2.00 | 13.50 |
| 100 | 5.60 | 53.51 |
| 200 | 16.86 | 254.10 |

## 6. Validation of the simulation program

The *g-functions* of the 4 × 4 bore field and of the 8 × 8 bore field obtained with 100 segments, illustrated in the previous section, were compared with those obtained by applying the Cimmino 2019 model [14], thanks to a Python code provided by the author using pygfunction [41]. The simulations were performed with the same number of segments and with the same 89 time instants in the range $-16 \le \ln(t^*) \le -6$.

The results of the comparison are shown in Fig. 8. The *g-functions* obtained by the two different models are graphically indistinguishable. The root-mean-square deviation between the two is 0.0010 for the $4 \times 4$ bore field and 0.0025 for the $8 \times 8$ bore field.

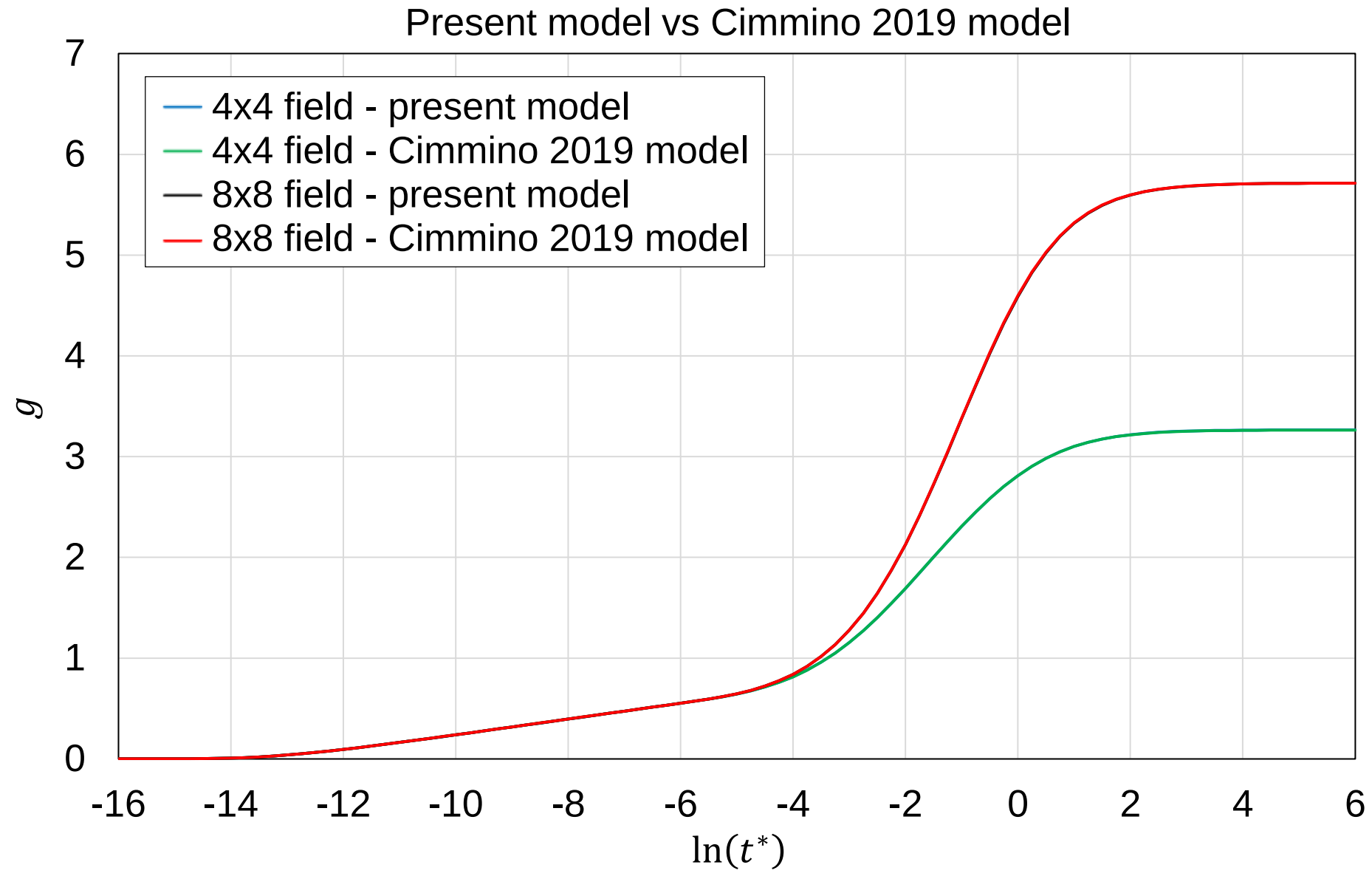

Figure 8: Comparison between the *g-functions* obtained by the present model and those obtained by the Cimmino 2019 model, for a $4 \times 4$ bore field and a $8 \times 8$ bore field with BHEs U94 1.6, $H_{\mathrm{b}} = 100$ m, $D = 1.8$ m, $B = 7.5$ m, $k_{\mathrm{g}} = 1.8$ W/(m K).

Another validation of the simulation program was performed, for the $4 \times 4$ bore field, by comparing the *g-function* and the *ftg-function* obtained by the present semi-analytical model with those obtained by finite-element simulations in COMSOL Multiphysics. Some details of the COMSOL simulations are reported below.

Since the field has two symmetry planes, only 4 BHEs were drawn in the computational domain. The latter was a parallelepiped having width and depth 263.75 m, namely 260 m plus $B/2$, and height 321.8 m, namely 320 m plus the buried depth, $D$. Each BHE was sketched by employing the conduction model proposed in Ref. [17]. Water was replaced by a solid cylinder with thermal conductivity $10^6$ W/(mK) and radius 20 mm, surrounded by a homogeneous cylindrical annulus with external radius equal to the BHE radius, representing the pipes and the grout. The thermal conductivity of this annulus was set to 2.0749 W/(mK), so that the thermal resistance of the annulus was equal to the asymptotic value of the 3D borehole thermal resistance, $R_{\mathrm{b3D}} = 0.1024$ m K/W (see Table 1). The volumetric heat capacity was set to 3.108 MJ/(m$^3$K) for both the central cylinder and the cylindrical annulus, so that the heat capacity per unit length of the model was equal to that of the real BHE. A uniform heat generation reproducing the heat rate supplied to the fluid was set in the solid cylinder representing water.

In order to have the same uniform temperature in all the cylinders representing water, these cylinders were interconnected by high-conductivity horizontal bars placed above the ground. The bars formed a high-conductivity square and were connected to the vertical cylinders representing water by high-conductivity cones. Both the bars and the cones had thermal conductivity $10^6\ \mathrm{W/(m\ K)}$ and volumetric heat capacity $0.01\ \mathrm{J/(m^3K)}$. An insulating vertical cylinder with zero thermal conductivity was placed between each cone and the ground, and an insulating disk with zero thermal conductivity was placed at the bottom of each BHE. As in Ref. [17], all the lengths in the vertical direction were reduced by a rescaling factor 20, and, as a consequence, the thermal conductivities in the vertical direction were reduced by a factor 400, for the ground and for the annuli representing the grout and the pipes.

The meshed computational domain is illustrated in Fig. 9. The image on the left shows the BHEs, the high-conductivity bars, and region of the computational domain close to the BHEs. The image on the right is a detail that illustrates the intersection of two high-conductivity bars and the connection between the bars and a vertical solid cylinder representing water (in transparency). The selected mesh is composed of 8,218,386 tetrahedral elements, and was obtained by setting maximum element size 2 m, minimum element size 0.01 m, maximum element growth rate 1.2, curvature factor 0.2, and resolution of narrow regions 1.

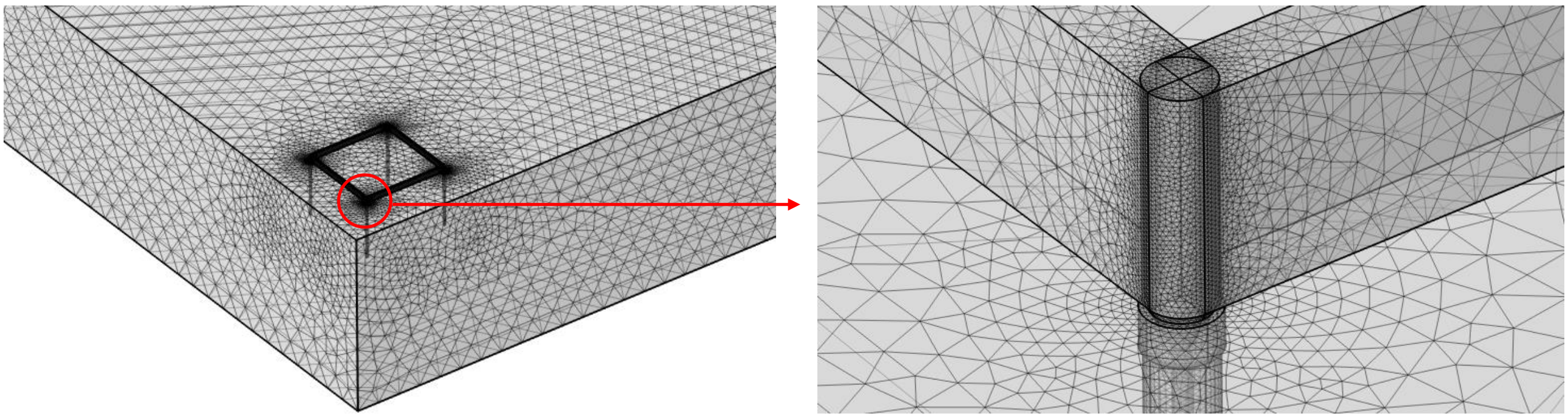

Figure 9: Portion of the computational domain, showing the vertical boundary planes close to the BHEs, the square of high-conductivity bars, and the BHEs (on the left), and detail showing the intersection of two horizontal bars and the connection between the bars and the high-conductivity solid representing water (on the right).

The ground surface, the horizontal bottom surface of the ground and the vertical boundaries of the ground far from the BHEs were set as isothermal, with temperature equal to zero. All the other boundaries of the computational domain were set as adiabatic. The mean heat rate supplied to the BHEs was set to $1\ \mathrm{W/m}$ . This way, the mean temperature of the external surface of the BHEs and that of the solid representing water, multiplied by 1.8, yield the *g-function* and the *ftg-function*, respectively.

The simulation was performed with time in seconds and outputs collected in the time interval from $\ln(t) = 0.339452$ to $\ln(t) = 27.339452$, with steps $\Delta\ln(t) = 0.25$. The initial and the final instant correspond to $\ln(t^*) = -21$ and $\ln(t^*) = 6$, respectively. The initial time step was set to 0.001 s. The other time steps employed in the computation were selected by the software, in order to fulfil the absolute tolerance 0.0001. The differential equation of transient heat conduction to be solved in each material, as well as the continuity of temperature and heat flux at the interface between different materials, were set by the software.

As for the semi-analytical model, the C++ program was run both in the option where it estimates a constant value of $R_{\mathrm{b3D}}$, and in the option where it reads the values of $R_{\mathrm{b3D}}(t_i^*)$ from an input file. In the second case, the values of $R_{\mathrm{b3D}}(t_i^*)$ employed were those calculated for the BHE U94 1.6 by the finite-element simulation described in Section 4.

The comparison between the *g-functions* obtained by the C++ program in the two options, and that obtained by the finite-element simulations is illustrated in Fig. 10, in the range $-16 \leq \ln(t^*) \leq -6$. The *g-functions* obtained by the C++ program in the two options are graphically indistinguishable from each other. Indeed, as it will be shown in the next section (see Fig. 12), the effect of $R_{\mathrm{b3D}}$ on the *g-function* is negligible for $\ln(t^*) < -4$. After that time, as shown in Fig. 5, $R_{\mathrm{b3D}}$ has already reached its asymptotic value, which is very close to the constant value of $R_{\mathrm{b3D}}$ estimated by the program.

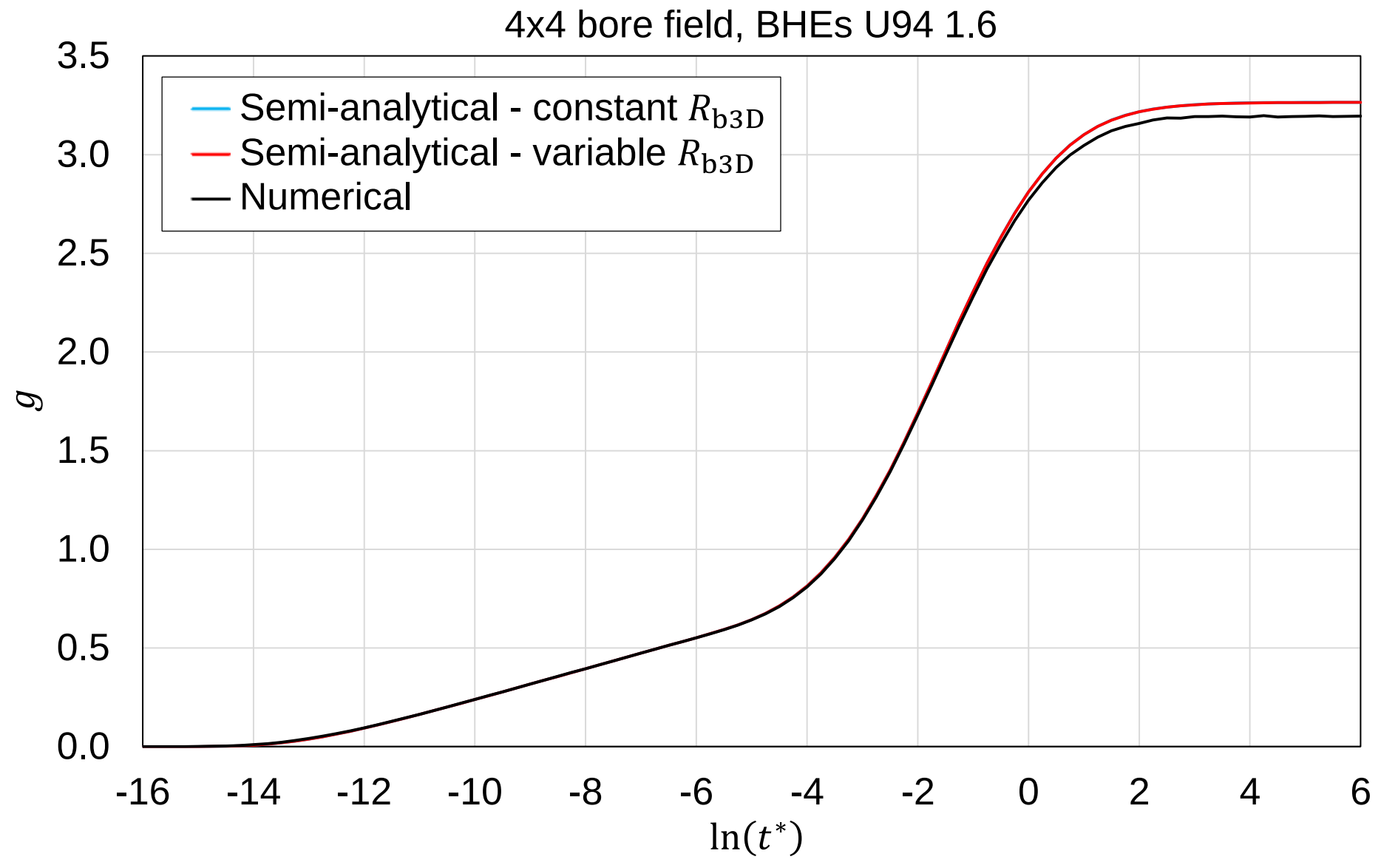

Figure 10: Comparison between the *g-function* obtained by the semi-analytical method with constant $R_{\mathrm{b3D}}$, the one obtained by the semi-analytical method with variable $R_{\mathrm{b3D}}$, and the one obtained by the finite-element simulation, for a $4 \times 4$ bore field with BHEs U94 1.6, $H_{\mathrm{b}} = 100$ m, $D = 1.8$ m, $B = 7.5$ m, $k_{\mathrm{g}} = 1.8$ W/(m K).

The agreement between the numerical *g-function* and the semi-analytical ones is excellent up to $\ln(t^*) = -1$, that corresponds to 21.6 years, and is still good up to $\ln(t^*) = 0$, that corresponds to 58.7 years. The root mean square difference between numerical *g-function* and the semi-analytical one computed with variable $R_{\text{b3D}}$ is 0.0062 in the range $-16 \le \ln(t^*) \le -1$, and 0.025 in the range $-1 \le \ln(t^*) \le 0$. The finite-element model, built to be accurate in the short and medium term, becomes less accurate in the very long term, namely after hundreds or thousands of years.

The comparison between the *ftg-functions* obtained by the C++ program in the two options and that obtained by the finite-element simulations is illustrated in Fig. 11, in the range $-20 \le \ln(t^*) \le -6$, i.e., from 3.8 s to 23,690 years. The semi-analytical *ftg-functions* obtained in the two options are graphically indistinguishable from each other for $\ln(t^*) > -12$, i.e., $t > 3.16$ hours. On the contrary, for very low values of $\ln(t^*)$, the semi-analytical *ftg-function* with constant $R_{\text{b3D}}$ is sharply greater than that with variable $R_{\text{b3D}}$. Indeed, the *ftg-function* is the sum of the *g-function* and $R^*_{\text{b3D}}$, and the constant value of $R_{\text{b3D}}$ estimated by the program is much greater than the true value, in the very short term.

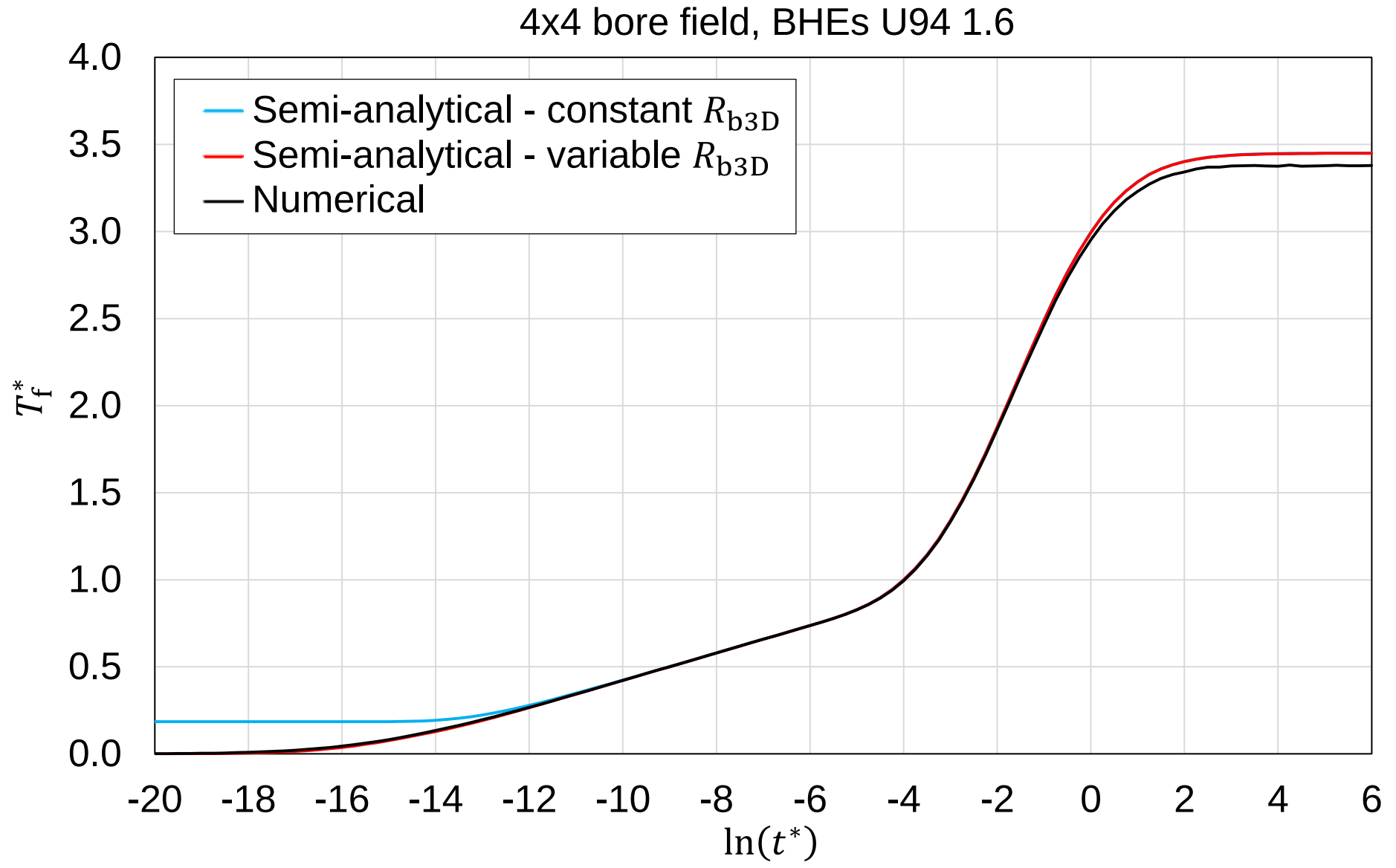

Figure 11: Comparison between the *ftg-function* obtained by the semi-analytical method with constant $R_{\text{b3D}}$, that obtained by the semi-analytical method with variable $R_{\text{b3D}}$, and that obtained by the finite-element simulation, for a $4 \times 4$ bore field with BHEs U94 1.6, $H_{\text{b}} = 100$ m, $D = 1.8$ m, $B = 7.5$ m, $k_{\text{g}} = 1.8$ W/(m K).

The agreement between the semi-analytical *ftg-function* computed with variable $R_{\text{b3D}}$ and the numerical *ftg-function* is excellent up to $\ln(t^*) = -1$, and is still good up to $\ln(t^*) = 0$. The root mean square difference between the two is 0.0064 in the range $-20 \le \ln(t^*) \le -1$, and 0.034 in the range $-1 \le \ln(t^*) \le 0$.

## 7. Inaccuracies caused by the assumptions of uniform heat rate and of uniform surface temperature of the BHEs

In this section, we illustrate the inaccuracies caused by the simplified assumptions of either uniform heat rate or uniform temperature of the surface of the BHEs, by comparing the *g-functions* obtained for different values of $R_{\mathrm{b3D}}$ for two square bore fields. The bore fields considered are a $4 \times 4$ bore field and a $10 \times 10$ bore field, with $H_{\mathrm{b}} = 100$ m, $D = 1.8$ m, $B = 7.5$ m, and $k_{\mathrm{g}} = 1.8\ \mathrm{W/(m\ K)}$. The simulations were performed for 5 constant values of $R_{\mathrm{b3D}}$: $0.001\ \mathrm{m\ K/W}$, $0.0632\ \mathrm{m\ K/W}$, $0.1030\ \mathrm{m\ K/W}$, $0.1808\ \mathrm{m\ K/W}$ and $1000\ \mathrm{m\ K/W}$. The second, the third and the fourth value are the program estimates of $R_{\mathrm{b3D}}$ for the BHEs 2U102 1.6, U94 1.6, and U54 1.0, respectively.

The *g-functions* obtained by the simulations of the $4 \times 4$ bore field are shown in Fig. 12. The figure shows that the effect of $R_{\mathrm{b3D}}$ on the *g-function* increases over time, becoming appreciable for $\ln(t^*) > -3$. In this range, the *g-function* increases with $R_{\mathrm{b3D}}$.

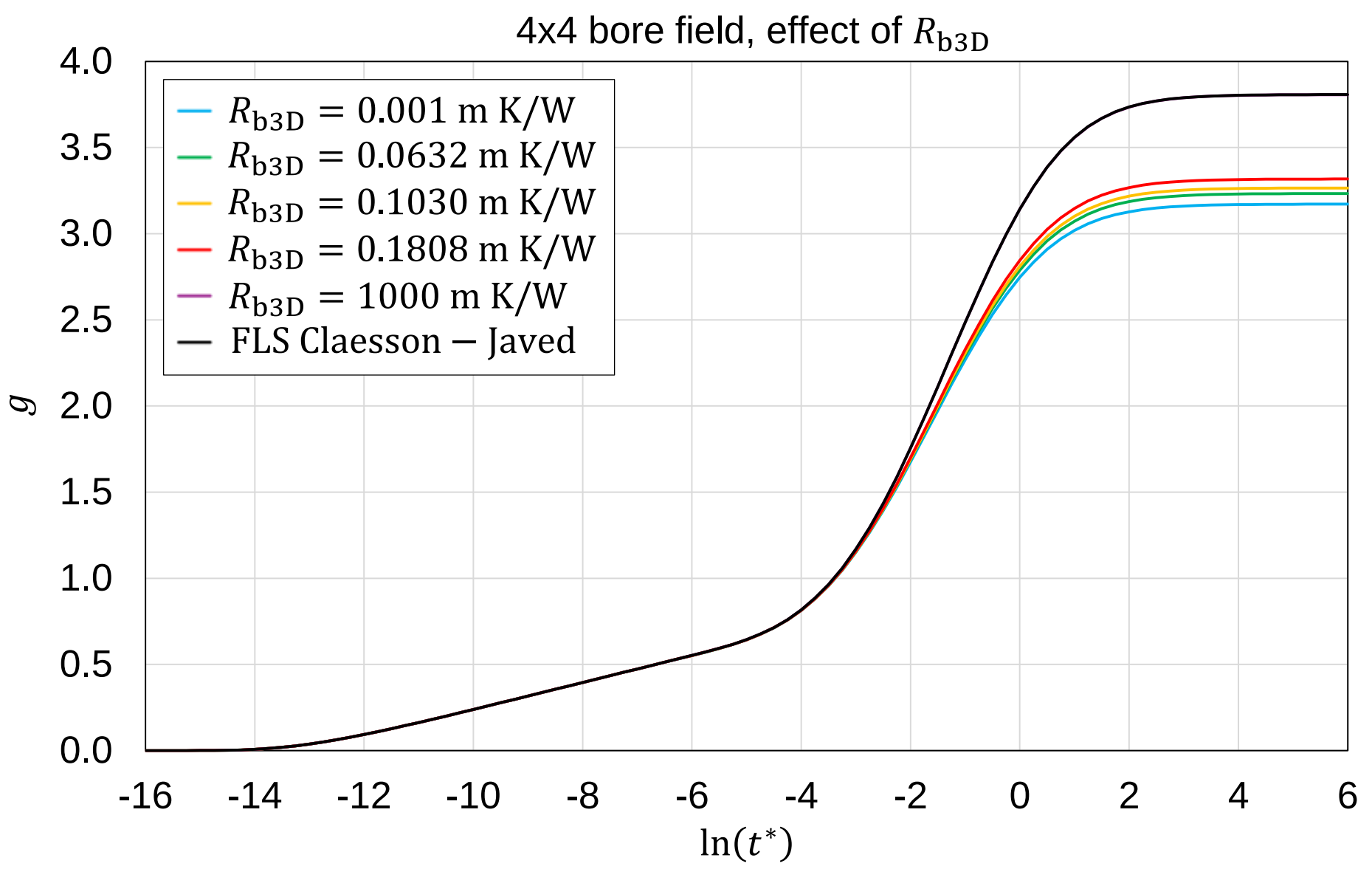

Figure 12: *g-functions* of a $4 \times 4$ bore field with $H_{\mathrm{b}} = 100$ m, $D = 1.8$ m, $B = 7.5$ m, $k_{\mathrm{g}} = 1.8\ \mathrm{W/(m\ K)}$, obtained for 5 different values of $R_{\mathrm{b3D}}$ and in the condition of uniform heat rate per unit length (FLS Claesson-Javed).

For $R_{\mathrm{b3D}} = 0.001\ \mathrm{m\ K/W}$, the temperature of the external surface of the BHEs is almost identical to the temperature of the fluid, which is set to be uniform. Therefore, the *g-function* obtained for $R_{\mathrm{b3D}} = 0.001\ \mathrm{mK/W}$ almost coincides with Eskilson's *g-function*, which is obtained in the condition of uniform external surface of the BHEs.

For $R_{\mathrm{b3D}} = 1000\ \mathrm{m\ K/W}$, the sums in Eq. (22) can be neglected, so that the heat loads of the segments are almost identical. Therefore, the *g-function* obtained for $R_{\mathrm{b3D}} = 1000\ \mathrm{mK/W}$ almost coincides with that yielded by the condition of uniform heat flux per unit length. This was proven by comparing the *g-function* for $R_{\mathrm{b3D}} = 1000\ \mathrm{mK/W}$ with that determined by applying the FLS solution of Claesson and Javed [6] for a single BHE with uniform heat flux and the superposition of the effects in space. As shown in Fig. 12, the two *g-functions* are graphically indistinguishable.

The *g-functions* of the BHEs 2U102 1.6, U94 1.6, and U54 1.0 are closer to the *g-function* yielded by the condition of uniform surface temperature, namely Eskilson's *g-function*, than to the one obtained by the FLS solution of Claesson and Javed. Therefore, the overestimation of the *g-function* caused by assuming a uniform heat load is greater than the underestimation made by employing the condition of uniform temperature of the external surface of the BHEs.

For the BHE U94 1.6, the overestimation caused by the assumption of uniform heat rate is 11.8% for $\ln(t^*) = 0$ and 16.6% for $\ln(t^*) = 6$, while the underestimation of the *g-function* caused by the assumption of uniform surface temperature is 2.27% for $\ln(t^*) = 0$ and 2.87% for $\ln(t^*) = 6$. The underestimation caused by the assumption of uniform surface temperature becomes higher for less performant BHEs. For the BHE U54 1.0, it is 3.47% for $\ln(t^*) = 0$ and 4.39% for $\ln(t^*) = 6$. Note that, for $H_b = 100$ m and $\alpha_g = 0.6 \times 10^{-6}$ m$^2$/s, $\ln(t^*) = 0$ corresponds to 58.7 years.

The inaccuracies become more significant for larger bore fields. This is shown in Fig. 13, which illustrates the *g-functions* obtained by the simulations of the $10 \times 10$ bore field.

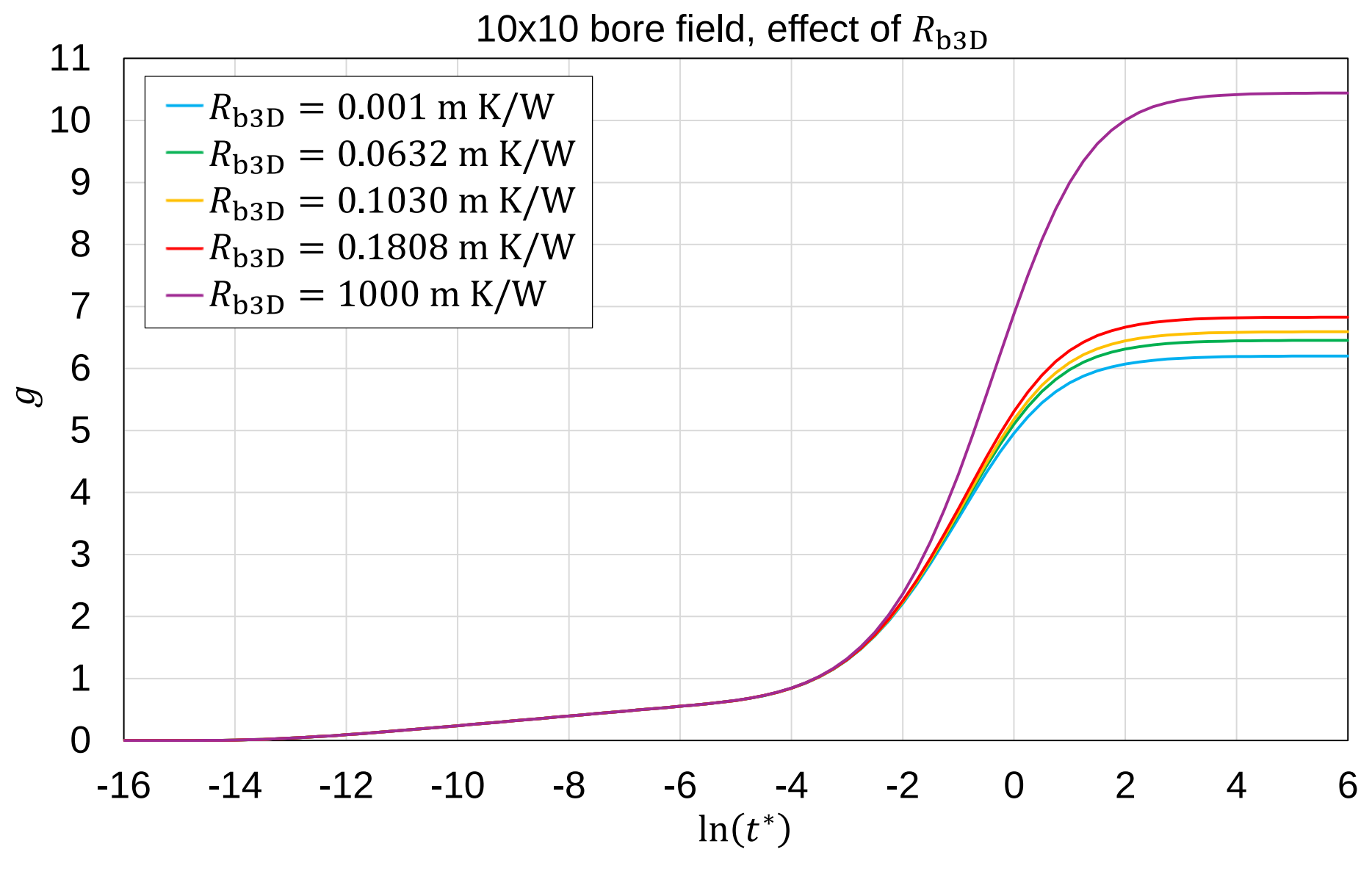

Figure 13: *g-functions* of a $10 \times 10$ bore field with $H_b = 100$ m, $D = 1.8$ m, $B = 7.5$ m, $k_g = 1.8$ W/(m K), obtained for 5 different values of $R_{b3D}$

In this case, for the BHE U94 1.6, the overestimation caused by the assumption of uniform heat rate is 32.9 % for $\ln(t^*) = 0$ and 58.4 % for $\ln(t^*) = 6$, while the underestimation of the *g-function* caused by the assumption of uniform surface temperature is 4.27 % for $\ln(t^*) = 0$ and 5.95 % for $\ln(t^*) = 6$. For the BHE U54 1.0, the underestimation of the *g-function* caused by the assumption of uniform surface temperature is 6.58 % for $\ln(t^*) = 0$ and 9.18 % for $\ln(t^*) = 6$.

## 8. Choice of the BHE length and of the geometry of the bore field, for a given plot of land

In this section, we show how the program can be employed to choose the optimal length of the BHEs and the best geometry of the bore field, for a given plot of land, kind of BHE, and total length of the bore field. Let us assume that the available plot of land is a square with sides of 60 m, the selected type of BHEs is U94 1.6 and the total length of the bore field to be installed is 8000 m. Let us also assume that the length of each BHE must be in the range $80\ \mathrm{m} \leq H_\mathrm{b} \leq 125\ \mathrm{m}$, and that, in this range, the undisturbed ground temperature is independent of $H_\mathrm{b}$, and the thermal conductivity and the thermal diffusivity of the ground are $k_\mathrm{g} = 1.8\ \mathrm{W/(m\ K)}$ and $\alpha_\mathrm{g} = 0.6 \times 10^{-6}\ \mathrm{m^2/s}$.

First, let us fix the geometry of the field, considering only bore fields of square shape. With this restriction, there are only three possibilities: one can either build a $10 \times 10$ bore field, with BHEs of length $H_\mathrm{b} = 80$ m spaced at a distance $B = 6.667$ m, or a $9 \times 9$ bore field, with BHEs of length $H_\mathrm{b} = 98.765$ m spaced at a distance $B = 7.5$ m, or an $8 \times 8$ bore field, with BHEs of length $H_\mathrm{b} = 125$ m spaced at a distance $B = 8.571$ m.

Since different values of $H_\mathrm{b}$ yield different ratios between $t$ and $t^*$, we calculated the *ftg-functions* for the same values of $t$, in the time range $1 \leq \log_{10}(t_\mathrm{hours}) \leq 8$ with a time step $\Delta\log_{10}(t_\mathrm{hours}) = 0.1$, where $t_\mathrm{hours}$ is the time expressed in hours. The simulations were performed with the constant values of $R_\mathrm{b3D}$ estimated by the program, namely $0.1017\ \mathrm{m\ K/W}$, $0.1029\ \mathrm{m\ K/W}$ and $0.1050\ \mathrm{m\ K/W}$, for $H_\mathrm{b} = 80$ m, $H_\mathrm{b} = 98.765$ m and $H_\mathrm{b} = 125$ m, respectively. The results are shown in Fig. 14.

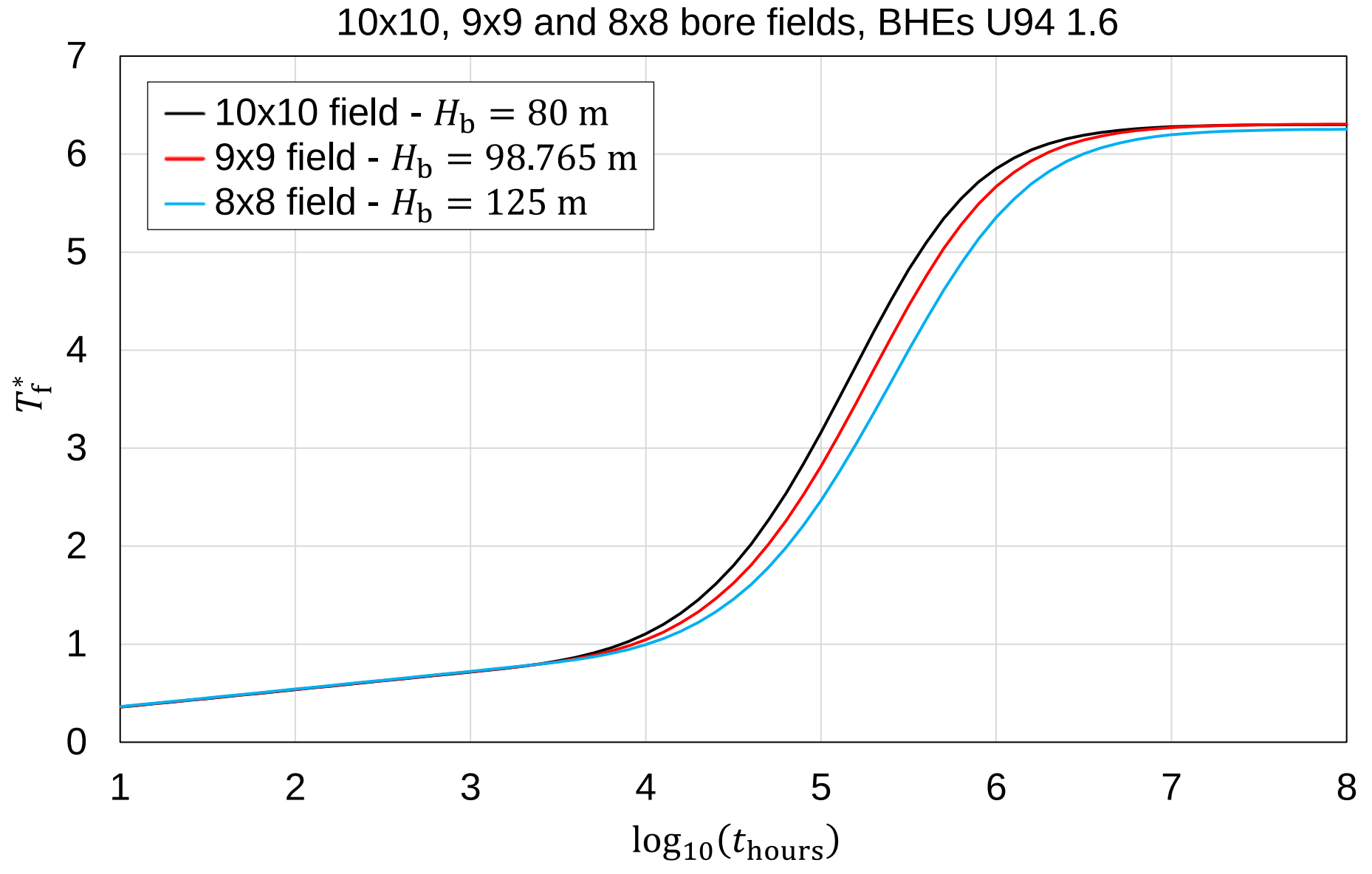

Figure 14: *ftg-functions* of a $10 \times 10$, a $9 \times 9$, and an $8 \times 8$ bore field, in a square plot of land with sides of 60 m, with BHEs U94 1.6, $D = 1.8$ m, $k_\mathrm{g} = 1.8\ \mathrm{W/(m\ K)}$ and $\alpha_\mathrm{g} = 0.6 \times 10^{-6}\ \mathrm{m^2/s}$.

The figure shows that the gap between the *ftg-functions* of the three fields is significant in the time range $4 \le \log_{10}(t_{\text{hours}}) \le 7$. In this range, the *ftg-function* of the $8 \times 8$ bore field has the lowest values, so the best choice for the BHE length is $H_{\text{b}} = 125$ m.

After selecting the BHE length, we checked whether the *ftg-function* can be further reduced by changing the geometry of the bore field. To do so, we first computed the mean dimensionless heat load of each BHE. We denoted this quantity as $\bar{a}_{uv}$, where $u$ and $v$ are, respectively, the row number and the column number of the BHE considered, ranging from 0 to 7. The time evolution of $\bar{a}_{uv}$ is illustrated, for non-equivalent BHEs, in Figure 15.

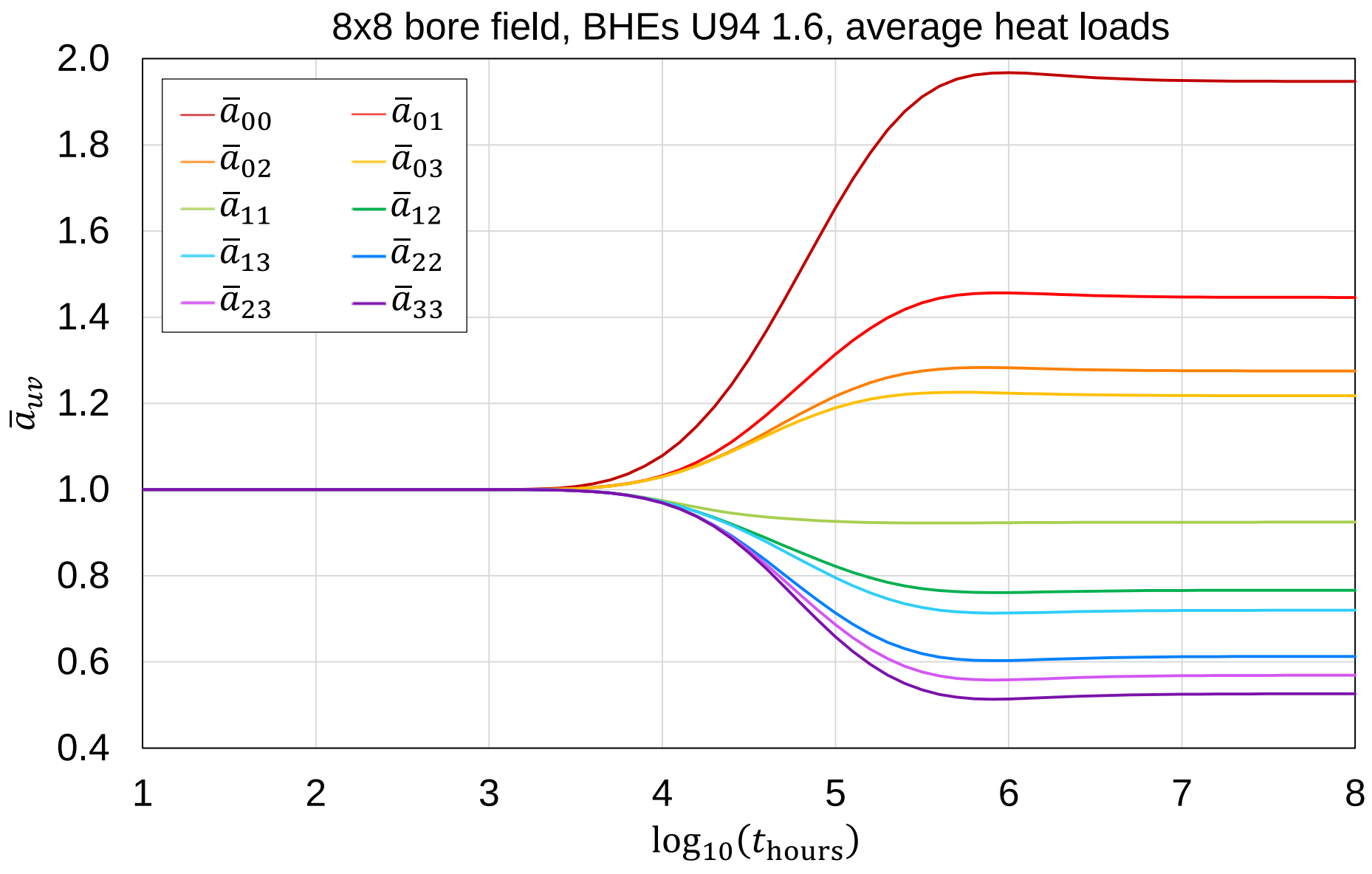

Figure 15: Mean dimensionless heat loads of the BHEs of the $8 \times 8$ bore field, with BHEs U94 1.6, $D = 1.8$ m, $k_{\text{g}} = 1.8$ W/(m K) and $\alpha_{\text{g}} = 0.6 \times 10^{-6}$ m$^2$/s.

The figure shows that $\bar{a}_{uv}$ becomes very low for the central BHEs for $\log_{10}(t_{\text{hours}}) > 5$, namely after 11 and a half years. As a consequence, it may be convenient to move the central BHEs towards the sides of the plot of land. Therefore, we considered three more bore fields, denoted as field a, field b, and field c, which are shown in Fig. 16.

Field a was obtained from the $8 \times 8$ bore field by displacing the four central BHEs to the sides of the plot of land. This field is composed of three concentric square rings at a distance of 8.571 m: the outer ring hosts 32 BHEs, the intermediate ring 20 and the inner ring 12. Field b was obtained from field a by displacing four BHEs from the inner ring to the outer ring. Since, in this configuration, the inner ring only hosts eight BHEs, the distance between the rings was increased to 10 m. Lastly, field c was obtained from field b by displacing four more BHEs from the inner to the outer ring.

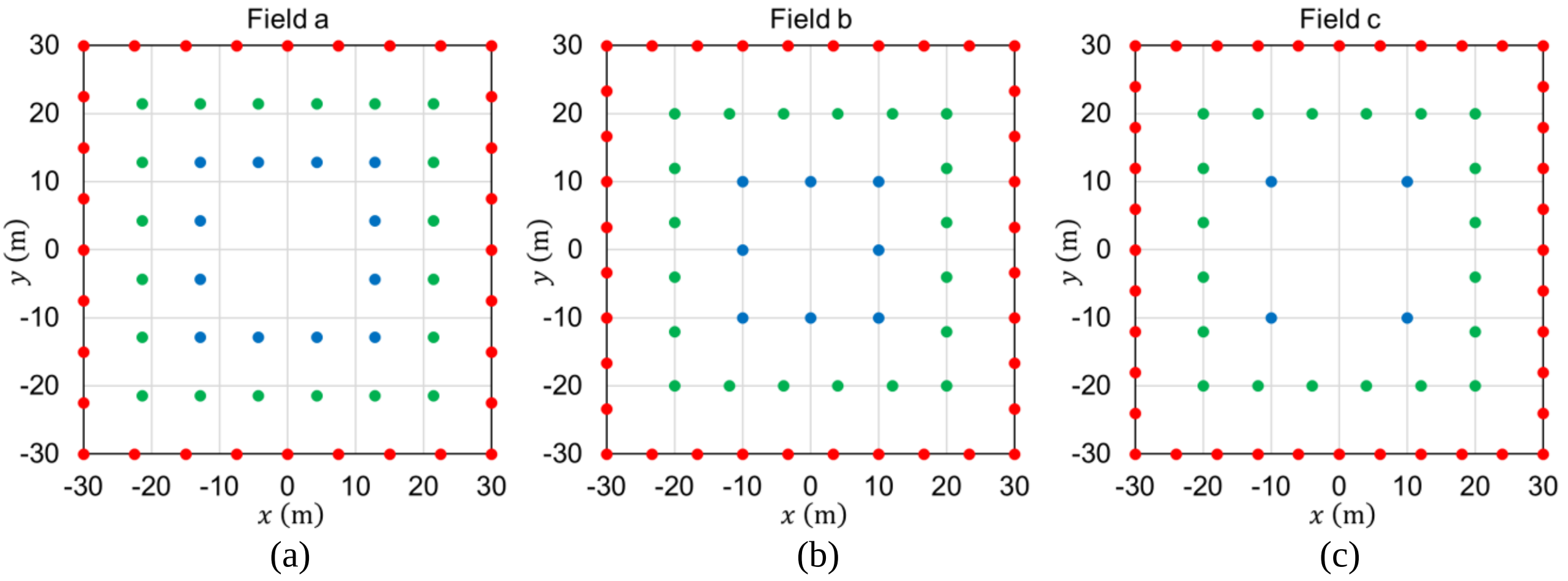

Figure 16: Alternatives to the 8 × 8 bore field. The BHEs are organized in three concentric square rings, and depicted in different colors according to the ring they belong to. The distance between the rings is 8.571 m in the first field (a), 10 m in the second field (b) and 10 m in the third field (c).

The *ftg-functions* of the 8 × 8 bore field, and of the fields a, b, and c are illustrated in Fig. 17. The differences are rather small, but all the alternatives to the 8 × 8 bore field perform better than it.

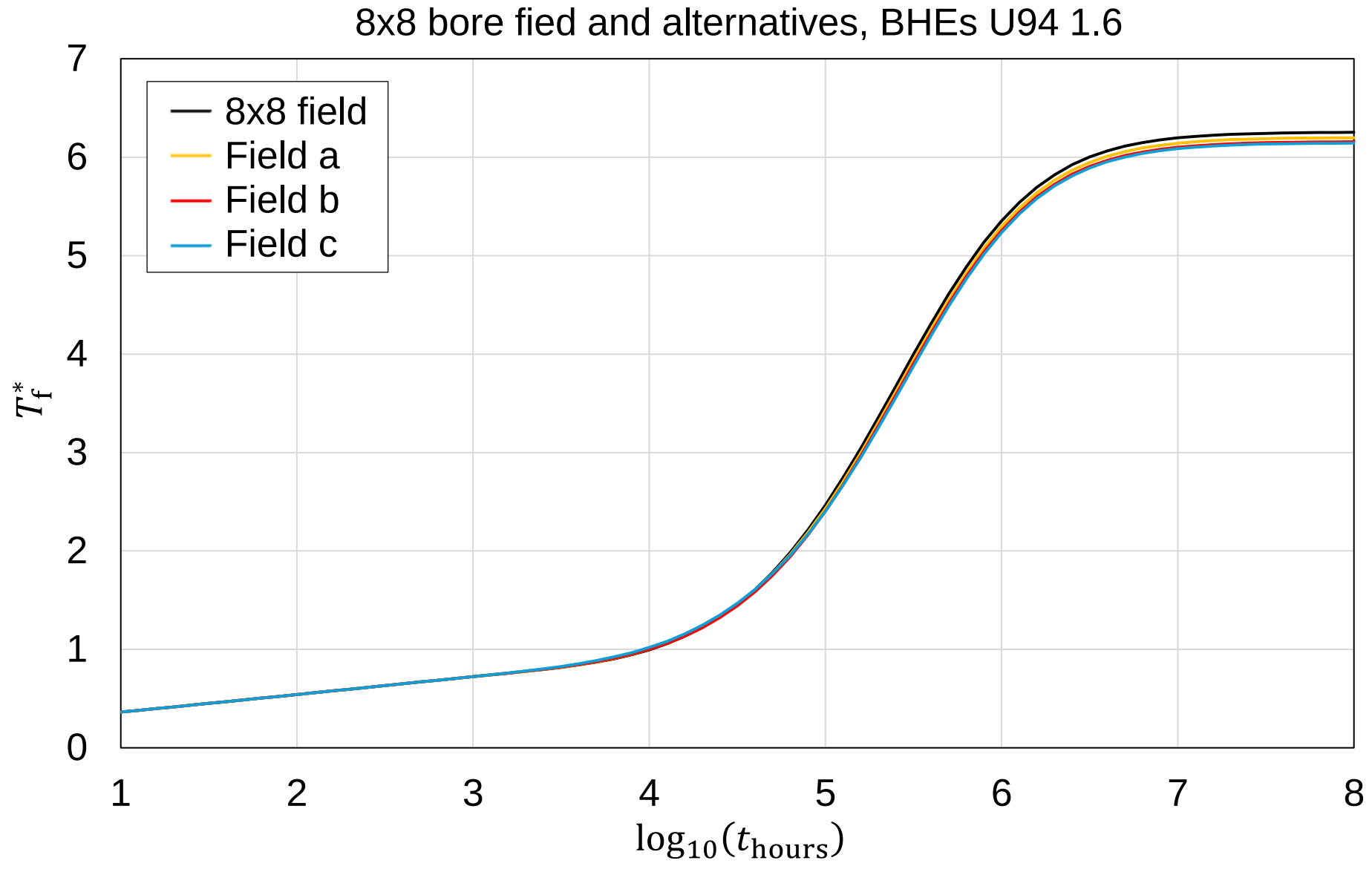

Figure 17: *ftg-functions* of the 8 × 8 bore field and of fields a, b, and c, placed in a square plot of land with sides of 60 m, with BHEs U94 1.6, $D = 1.8$ m, $k_g = 1.8$ W/(m K) and $\alpha_g = 0.6 \times 10^{-6}$ m$^2$/s.

A clearer comparison is illustrated in Fig. 18, which reports the difference between the *ftg-function* of fields a, b and c, and the *ftg-function* of the 8 × 8 bore field. The time axis is in linear scale, in years ($t_{\text{years}}$), and covers the first 60 years, namely the time period in which the bore field is presumably in use.

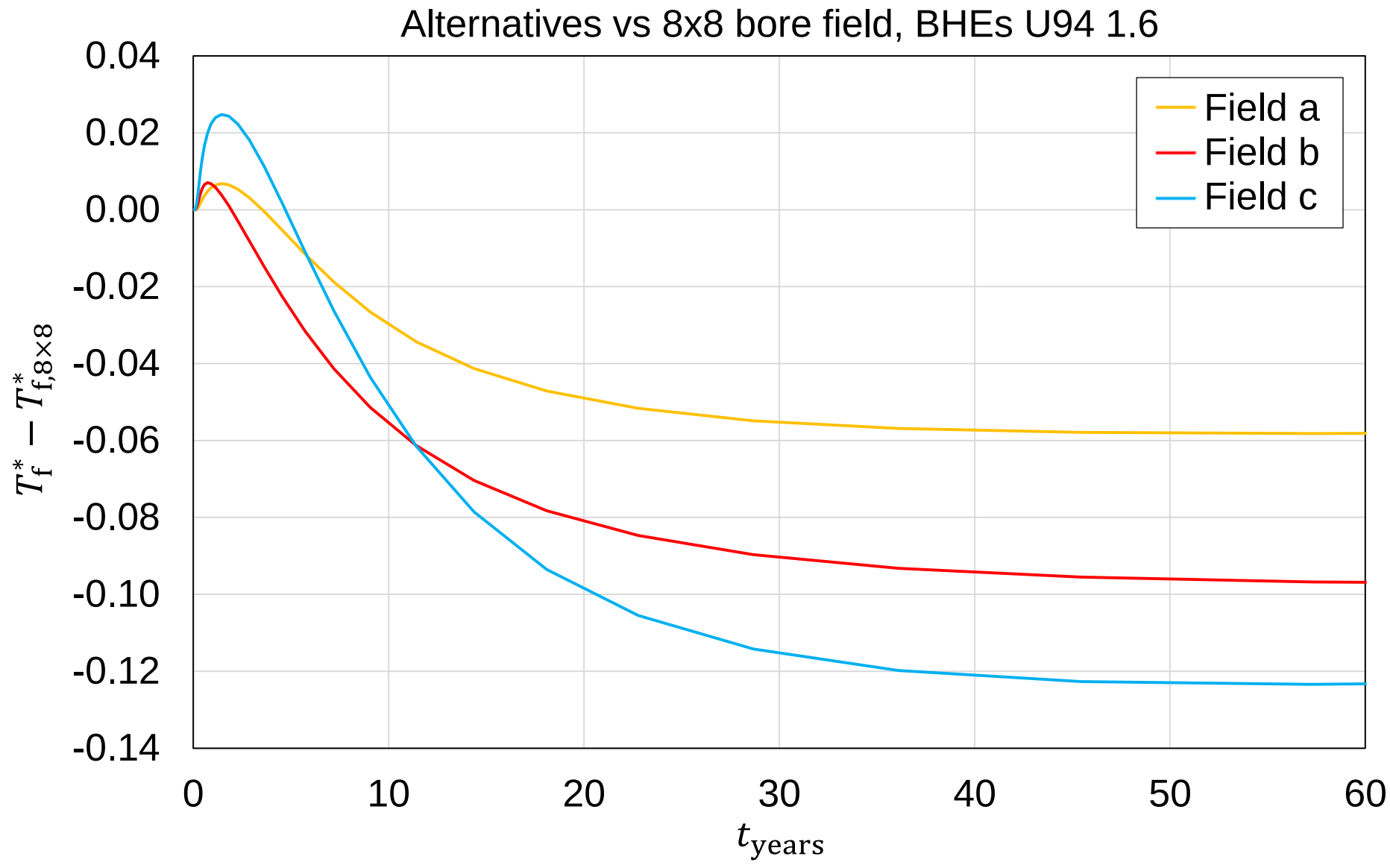

Figure 18: Differences between the *ftg-functions* of fields a, b, and c, and the *ftg-function* of the $8 \times 8$ bore field, for a square plot of land with sides of 60 m, with BHEs U94 1.6, $D = 1.8$ m, $k_g = 1.8$ W/(m K) and $\alpha_g = 0.6 \times 10^{-6}$ m$^2$/s.

The figure shows that the best performance is obtained by fields b and c. Field b is slightly worse than the $8 \times 8$ bore field only for the first two years, then it becomes significantly better. Field c is worse than the $8 \times 8$ bore field for the first four and a half years, but it quickly gets better, becoming the most efficient one after 11 and a half years.

## 9. Conclusions

A fast and accurate semi-analytical method has been developed to determine the thermal response of borehole-heat-exchanger (BHE) fields. The method overcomes the inaccuracies of the ones usually employed, which are based either on the condition of uniform heat rate per unit length or on the condition of uniform temperature of the external surface of the BHEs. It is a simpler and equally precise alternative to the excellent method proposed by Cimmino (Int J Heat Mass Tran 91, 2015).
The method applies to any bore field with BHEs having the same length, buried at the same depth, and fed in parallel with an equal inlet fluid temperature. It can be used to determine both the *g-function* of a bore field, namely the time evolution of the dimensionless mean temperature of the external surface of the BHEs, and the *ftg-function*, namely the time evolution of the dimensionless mean temperature of the fluid.
Each BHE is divided into segments. The interaction between the segments is calculated by applying the finite line-source solutions for BHE segments determined by Cimmino and Bernier (Int J Heat Mass Tran 70, 2014) and the superposition of the effects in space and time. Each segment is subjected to a uniform time dependent heat load per unit length, which is determined by solving a linear system of equations that impose the conditions of uniform fluid temperature in the bore field and constant total heat load. The relation between the fluid temperature and the mean temperature of the external surface of the BHEs is set by introducing a 3D borehole thermal resistance, $R_{\mathrm{b3D}}$, that considers the thermal short-circuiting between the descending and the ascending flow. This way, the condition of uniform fluid temperature yields the same *g-function* and *ftg-function* as a model involving the energy balance along the flow.

The method has been implemented in a C++ program, available at the open-source online data repository of the University of Bologna. The program yields the *g-function*, the *ftg-function* and the time evolution of the dimensionless heat load of each BHE segment. The finite line-source solutions are computed by numerical integrals with the trapezoidal rule, and the linear systems of equations are solved by the Gaussian elimination method. Thanks to the optimizations introduced in the computations of the finite line-source solutions and in the solution of the linear systems, the execution time of the program is extremely low, as shown by the examples in Table 2.

The program can be run with two different options: it can either estimate a constant value of $R_{\mathrm{b3D}}$ based on the characteristics of the BHEs and on the properties of the fluid, or it can read the value of $R_{\mathrm{b3D}}$ at each time instant from an input file supplied by the user. In the first option, the program yields a very accurate *g-function* in a time range from a few minutes to thousands of years, and a reasonably accurate *ftg-function* from a few hours to thousands of years. In the second option, if the supplied values of $R_{\mathrm{b3D}}$ are precise, the program yields both an accurate *g-function* and an accurate *ftg-function* from a few seconds to thousands of years.

The program has been employed to determine the overestimation of the *g-function* caused by the assumption of uniform heat rate and the underestimation of the *g-function* caused by the assumption of uniform surface temperature of the BHEs. The results have shown that the condition of uniform surface temperature yields a much more accurate *g-function* than the condition of uniform heat rate. Even the latter, however, produces non-negligible errors in the *g-function*, which can reach up to 10% for large bore fields, in the long term.

The program has also been used to illustrate the low performance of the central BHEs in large and compact bore fields, and to show how the bore field can be optimized, for a given plot of land and a fixed total length of the field.

**Acknowledgements**

The authors are grateful to Professor Massimo Cimmino for providing the Python code that was used for the validation of the simulation program. This work was supported by National Recovery and Resilience Plan (NRRP), Mission 4 Component 2 Investment 1.5 - Call for tender No. 3277 of 30/12/2021 of Italian Ministry of University and Research funded by the European Union – NextGenerationEU; project code ECS00000033, Concession Decree No. 1052 of 23/06/2022 adopted by the Italian Ministry of University and Research, CUP D93C22000460001, “Ecosystem for Sustainable Transition in Emilia- Romagna” (Ecosister), Spoke 4.

**References**

1. M. Rivoire, A. Casasso, B. Piga, R. Sethi, R. Assessment of Energetic, Economic and Environmental Performance of Ground-Coupled Heat Pumps, Energies 11 (2018) 1941, https://doi.org/10.3390/en11081941
2. P. Eskilson, Thermal analysis of heat extraction boreholes, Ph.D. Thesis, University of Lund, Lund, Sweden, 1987.
3. H.Y. Zeng, N.R. Diao, Z.H. Fang, A finite line-source model for boreholes in geothermal heat exchangers, *Heat Transfer–Asian Research* 31 (2002) 558–567, https://doi.org/10.1002/htj.10057
4. L. Lamarche, B. Beauchamp, A new contribution to the finite line-source model for geothermal boreholes, *Energ. Buildings* 39 (2007) 188–198, https://doi.org/10.1016/j.enbuild.2006.06.003
5. T.V. Bandos, A. Montero, E. Fernandez, J.L.G. Santander, J.M. Isidro, J. Perez, P.J. Fernandez de Cordoba, J.F. Urchueguía, Finite line-source model for borehole heat exchangers: effect of vertical temperature variations, Geothermics 38 (2009) 263–270, https://doi.org/10.1016/j.geothermics.2009.01.003
6. J. Claesson, S. Javed, An analytical method to calculate borehole fluid temperatures for time-scales from minutes to decades, *ASHRAE Tran* 117(2) (2011) 279–288.
7. M. Cimmino, M. Bernier, A semi-analytical method to generate g-functions for geothermal bore fields, *Int J Heat Mass Tran* 70 (2014) 641–650, https://doi.org/10.1016/j.ijheatmasstransfer.2013.11.037
8. L. Lamarche, G-function generation using a piecewise-linear profile applied to ground heat exchangers, Int. J. Heat Mass Transfer 115 (2017) 354–360, https://doi.org/10.1016/j.ijheatmasstransfer.2017.08.051
9. M. Cimmino, Fast calculation of the g-functions of geothermal borehole fields using similarities in the evaluation of the finite line source solution. J. Build. Perform. Simul. 11 (2018) 655–668, https://doi.org/10.1080/19401493.2017.1423390
10. P. Monzó, P. Mogensen, J. Acuña, F. Ruiz-Calvo, C. Montagud, A novel numerical approach for imposing a temperature boundary condition at the borehole wall in borehole fields, Geothermics 56 (2015) 35–44. https://doi.org/10.1016/j.geothermics.2015.03.003

11. C. Naldi, E. Zanchini, A new numerical method to determine isothermal g-functions of borehole heat exchanger fields, Geothermics 77 (2019) 278–287, https://doi.org/10.1016/j.geothermics.2018.10.007
12. P. Monzó, A.R. Puttige, J. Acuña, P. Mogensen, A. Cazorla, J. Rodriguez, C. Montagud, F. Cerdeira, Numerical modeling of ground thermal response with borehole heat exchangers connected in parallel, Energ Buildings 172 (2018) 371–384, https://doi.org/10.1016/j.enbuild.2018.04.057
13. M. Cimmino, The effects of borehole thermal resistances and fluid flow rate on the *g-functions* of geothermal bore fields, Int J Heat Mass Tran 91 (2015) 1119–1127, https://doi.org/10.1016/j.ijheatmasstransfer.2015.08.041
14. M. Cimmino, Semi-Analytical Method for g-Function Calculation of bore fields with series- and parallel-connected boreholes, Science and Technology for the Built Environment 25:8 (2019) 1007–1022, https://doi.org/10.1080/23744731.2019.1622937
15. A. Jahanbin, C. Naldi, E. Zanchini, Relation Between Mean Fluid Temperature and Outlet Temperature for Single U-Tube Boreholes, Energies 13 (2020) 828, https://doi:10.3390/en13040828
16. E. Zanchini, A. Jahanbin, Simple equations to evaluate the mean fluid temperature of double-U-tube borehole heat exchangers, Applied Energy 231 (2018) 320–330, https://doi.org/10.1016/j.apenergy.2018.09.094
17. E. Zanchini, Comparison between conduction models and models including the energy balance along the flow, for the simulation of U-tube borehole heat exchangers, Applied Thermal Engineering 257 (2024) 124311, https://doi.org/10.1016/j.applthermaleng.2024.124311
18. A. Zarrella, M. Scarpa, M. De Carli, Short time step analysis of vertical ground-coupled heat exchangers: The approach of CaRM, *Renew Energ* 36 (2011) 2357–2367, https://doi.org/10.1016/j.renene.2011.01.032
19. D. Bauer, W. Heidemann, H.-J.G. Diersch, Transient 3D analysis of borehole heat exchanger modeling, Geothermics 40 (2011) 250–260, https://doi.org10.1016/j.geothermics.2011.08.001
20. P. Pasquier, D. Marcotte, Short-term simulation of ground heat exchanger with an improved TRCM, *Renew Energ* 46 (2012) 92–99, https://doi.org/10.1016/j.renene.2012.03.014
21. F. Ruiz-Calvo, M. De Rosa, J. Acuña, J.M. Corberán, C. Montagud, Experimental validation of a short-term Borehole-to-Ground (B2G) dynamic model, *Appl Energ* 140 (2015) 210–223, https://doi.org/10.1016/j.apenergy.2014.12.002

22. Lamarche, L.; Beauchamp, B. New solutions for the short-time analysis of geothermal vertical boreholes. *Int J Heat Mass Tran* **2007**, *50*, 1408–1419. https://doi.org/10.1016/j.ijheatmasstransfer.2006.09.007
23. G. Bandyopadhyay, M. Kulkarni, M. Mann, A new approach to modeling ground heat exchangers in the initial phase of heat flux build-up, ASHRAE Transactions 114(2) (2008) 428–439.
24. G. Bandyopadhyay, W. Gosnold, M. Mann, Analytical and semi-analytical solutions for short-time transient response of ground heat exchangers, Energy and Buildings 40 (2008) 1816–1824, https://doi.org/10.1016/j.enbuild.2008.04.005
25. L. Lamarche, Short-time analysis of vertical boreholes, new analytic solutions and choice of equivalent radius, *Int J Heat Mass Tran* 91 (2015) 800–807, https://doi.org/10.1016/j.ijheatmasstransfer.2015.07.135
26. Y. Man, H. Yang, N. Diao, J. Liu, Z. Fang, A new model and analytical solutions for borehole and pile ground heat exchangers, *Int J Heat Mass Tran* 53 (2010) 2593–2601, https://doi.org/10.1016/j.ijheatmasstransfer.2010.03.001
27. M. Li, P. Li, V. Chan, A.C.K. Lai, Full-scale temperature response function (G-function) for heat transfer by borehole ground heat exchangers (GHEs) from sub-hour to decades, Applied Energy 136 (2014) 197–205, http://dx.doi.org/10.1016/j.apenergy.2014.09.013
28. E. Zanchini, C. Naldi, M. Dongellini, Dimensionless fluid-to-ground thermal response of single-line bore fields with isothermal fluid, Applied Thermal Engineering 233 (2023) 121210, https://doi.org/10.1016/j.applthermaleng.2023.121210
29. Z. Li, M. Zheng, Development of a numerical model for the simulation of vertical U-tube ground heat exchangers, *Applied Thermal Engineering* 29 (2009) 920–924, https://doi.10.1016/j.applthermaleng.2008.04.024
30. G.A. Florides, P. Christodoulides, P. Pouloupatis, An analysis of heat flow through a borehole heat exchanger validated model, Applied Energy 92 (2012) 523–533, https://doi.10.1016/j.apenergy.2011.11.064
31. S.J. Rees, M. He, A three-dimensional numerical model of borehole heat exchanger heat transfer and fluid flow, Geothermics 46 (2013) 1–13, http://dx.doi.org/10.1016/j.geothermics.2012.10.004
32. T.Y. Ozudogru, C.G. Olgun, A. Senol, 3D numerical modeling of vertical geothermal heat exchangers, Geothermics 51 (2014) 312–324, http://dx.doi.org/10.1016/j.geothermics.2014.02.005

33. X. Lei, X. Zheng, C. Duan, J. Ye, K. Liu, Three-Dimensional Numerical Simulation of Geothermal Field of Buried Pipe Group Coupled with Heat and Permeable Groundwater, Energies 12 (2019) 3698, https://doi:10.3390/en12193698
34. T. Amanzholov, Ye Belyayev, M. Mohanraj, A. Toleukhanov, Study of Borehole Heat Exchanger Heat Transfer Enhancement Parameters, Journal of Mathematics, Mechanics and Computer Science, Section 2, Mechanics 115 (2022) 3, https://doi.org/10.26577/JMMCS.2022.v115.i3.07
35. G. Hellström, Ground Heat Storage: Thermal Analysis of Duct Storage Systems. PhD Thesis, Department of Mathematical Physics, University of Lund, Lund, Sweden, 1991.
36. S. W. Churchill, Comprehensive correlating equations for heat, mass and momentum transfer in fully developed flow in smooth tubes, Ind. Eng. Chem. Fundam. 16 (1977) 109–116, https://doi.org/10.1021/i160061a021
37. L. Lamarche, S. Kajl, B. Beauchamp, A review of methods to evaluate borehole thermal resistances in geothermal heat-pump systems, Geothermics 39 (2010) 187–200, https://doi.org/10.1016/j.geothermics.2010.03.003
38. H. Zeng, N. Diao, Z. Fang, Heat transfer analysis of boreholes in vertical ground heat exchangers, *Int J Heat Mass Tran* 46 (2003) 4467–4481,
39. E. Zanchini, A. Jahanbin, Effects of the temperature distribution on the thermal resistance of double U-tube borehole heat exchangers, Geothermics 71 (2018) 46–54, http://dx.doi.org/10.1016/j.geothermics.2017.07.009
40. NIST Chemistry WebBook. Available online: https://webbook.nist.gov/chemistry/fluid
41. M. Cimmino, J.C. Cook, pygfunction 2.2: New features and improvements in accuracy and computational efficiency. In Research Conference Proceedings, IGSHPA Annual Conference 2022, International Ground Source Heat Pump Association, 2022, pp. 45–52. doi: https://doi.org/10.22488/okstate.22.000015